\documentclass[journal]{IEEEtran}
\IEEEoverridecommandlockouts
\usepackage{cite}
\usepackage{amsmath,amssymb,amsfonts}
\usepackage{graphicx}
\usepackage{textcomp}
\usepackage{xcolor}
\usepackage{algorithm}
\usepackage{algorithmic}
\usepackage{enumitem}
\usepackage{footnote}
\usepackage{multirow}
\usepackage{threeparttable}
\usepackage{makecell}
\usepackage[colorlinks,linkcolor=blue]{hyperref}

\makeatletter
\let\MYcaption\@makecaption
\makeatother

\usepackage[font=footnotesize]{subcaption}

\makeatletter
\let\@makecaption\MYcaption
\makeatother

\usepackage{array}

\def\BibTeX{{\rm B\kern-.05em{\sc i\kern-.025em b}\kern-.08em
    T\kern-.1667em\lower.7ex\hbox{E}\kern-.125emX}}

\usepackage{amsthm}

\theoremstyle{definition}
\newtheorem{definition}{Definition}

\makeatletter
\renewcommand{\maketag@@@}[1]{\hbox{\m@th\normalsize\normalfont#1}}%
\makeatother

\begin{document}

\title{Beyond-Diagonal RIS Under Non-Idealities: Learning-Based Architecture Discovery and Optimization \vspace{0.2em}}

\author{Binggui Zhou,~\IEEEmembership{Member,~IEEE},
        and Bruno Clerckx,~\IEEEmembership{Fellow,~IEEE}\vspace{-2em}
\thanks{This work has been partially supported by UKRI grant EP/X040569/1, EP/Y037197/1, EP/X04047X/1, EP/Y037243/1. (Corresponding author: Bruno Clerckx.)}
\thanks{
Binggui Zhou and Bruno Clerckx are with the Department of Electrical and Electronic Engineering, Imperial College London, London, SW7 2AZ, U.K. Bruno Clerckx is also with the Department of Electronic Engineering, Kyung Hee University, Yongin-si, Gyeonggi-do 17104, South Korea (email: {binggui.zhou, b.clerckx}@imperial.ac.uk).}
}
\maketitle

\begin{abstract}Beyond-diagonal reconfigurable intelligent surface (BD-RIS) has recently been introduced to enable advanced control over electromagnetic waves to further increase the benefits of traditional RIS in enhancing signal quality and improving spectral and energy efficiency for next-generation wireless networks. A significant issue in designing and deploying BD-RIS is the tradeoff between its performance and circuit complexity. While existing studies have explored optimal architectures to minimize circuit complexity in ideal BD-RIS, architecture discovery for non-ideal BD-RIS remains uninvestigated. Consequently, how non-idealities and circuit complexity jointly affect the performance of BD-RIS remains unclear, making it difficult to achieve the performance-circuit complexity tradeoff in the presence of non-idealities. Essentially, architecture discovery for non-ideal BD-RIS faces challenges from both the computational complexity of global architecture search and the difficulty in achieving global optima. To tackle these challenges, we propose a learning-based two-tier architecture discovery framework (LTTADF) consisting of an architecture generator and a performance optimizer to jointly discover optimal architectures for non-ideal BD-RIS given specific circuit complexities, which can effectively explore over a large architecture space while avoiding getting trapped in poor local optima and thus achieving near-optimal solutions for the performance optimization. Numerical results provide valuable insights for deploying non-ideal BD-RIS considering the performance-circuit complexity tradeoff. Specifically, mutual coupling at the BD-RIS does not affect the optimal BD-RIS architectures in MU-MIMO systems; increasing the circuit complexity of lossy BD-RIS can be detrimental, highlighting the necessity to learn an optimal architecture that balances circuit complexity and loss mitigation; the circuit complexity and quantization resolution of discrete-value BD-RIS can compensate for the limitations of each other.

\end{abstract}


\section{Introduction}

Reconfigurable intelligent surface (RIS) is regarded as a promising technology to meet the increasing demands for spectrum and energy efficiency in future wireless communication systems\cite{huang2019reconfigurable, wu2019intelligent}. RIS enables dynamic control over the signal propagation environment by shaping the trajectory and characteristics of electromagnetic waves, thereby opening new possibilities to enhance signal quality, suppress interference, extend coverage, and improve spectral and energy efficiency. Traditional RIS manipulates the propagation environment by adjusting its diagonal phase shift matrix, whereas beyond-diagonal RIS (BD-RIS) introduces tunable interconnections between RIS elements, and is hence not restricted to a diagonal scattering matrix, to enable advanced control over electromagnetic waves, which is expected to further enhance beamforming, interference mitigation, and coverage extension compared with traditional RIS\cite{shen2022modeling,li2024reconfigurable,li2026tutorial}.\footnote{Note that throughout this work, we focus exclusively on purely passive BD-RIS. Unlike recently proposed active RIS designs \cite{shen2026active, zhang2023active} that integrate power amplifiers to actively boost incident signals, our considered BD-RIS relies entirely on passive reconfigurable components, ensuring the fundamental low-power and low-cost advantages inherent to traditional RIS technologies.} Such benefits have been widely explored by existing works such as \cite{li2022reconfigurablea,li2023diagonal, nerini2024closedform, liu2026nonreciprocal, nerini2025localized,peng2026joint}. The capacity and coverage enhancements provided by BD-RIS are expected to serve as a critical physical-layer foundation for next-generation delay-sensitive and data-intensive applications.

Despite the great benefits offered by BD-RIS, a significant issue in designing and deploying BD-RIS is the tradeoff between its performance and circuit complexity \cite{nerini2023pareto}.\footnote{For simplicity, we only discuss the circuit complexity of reciprocal BD-RIS, and due to the symmetry of reciprocal networks, we define the circuit complexity as the number of independent tunable admittance components (i.e., the non-zero elements in the upper triangular part and the diagonal of the admittance matrix) in this paper.} Generally, the fully-connected BD-RIS, whose elements are all interconnected via tunable admittance components, has the highest circuit complexity $\frac{N_I(N_I+1)}{2}$ with $N_I$ being the number of RIS elements and thus can generally achieve the optimal performance. It is notable that the circuit complexity of fully-connected BD-RIS scales quadratically with $N_I$, which is unaffordable for large-dimensional BD-RIS. While the single-connected BD-RIS, where RIS elements are not interconnected, has the lowest circuit complexity $N_I$ and thus significantly underperforms the fully-connected BD-RIS. To balance the performance and circuit complexity, some recent works have explored optimal BD-RIS architectures with much lower circuit complexity. In \cite{nerini2024diagonal}, tree-connected BD-RIS was found to be as optimal as the fully-connected BD-RIS in single-user multiple-input single-output (SU-MISO) systems with only a circuit complexity of $2N_I-1$. In \cite{zhou2024novel} and \cite{wu2025beyonddiagonal}, two novel BD-RIS architectures, called stem-connected BD-RIS and band-connected BD-RIS, were proposed and demonstrated to be optimal for multi-user multiple-input multiple-output (MU-MIMO) cases while having a considerably low circuit complexity of $L(2N_I-2L+1)$ (where $L=\min\{N_R, N_T, \frac{N_I}{2}\}$ with $N_T$ being the number of transmit antennas and $N_k$ being the number of the $k$-th user's antennas). However, these works only considered ideal BD-RIS, neglecting non-idealities in practical BD-RIS, e.g., mutual coupling, losses in tunable admittance components, and quantization errors in the discrete-value admittance matrix. To explore the impact of mutual coupling, a global optimal closed-form solution for tree-connected BD-RIS with mutual coupling to maximize the channel gain in a single-user single-input single-output (SU-SISO) system was provided in \cite{nerini2024global}, demonstrating that the tree-connected BD-RIS architecture was still the optimal BD-RIS architecture for BD-RIS with mutual coupling in SU-SISO. In addition, performance optimization for lossy BD-RIS in SU-SISO and multi-user MISO (MU-MISO) systems was presented in \cite{peng2026lossy}. Moreover, the impact of quantization errors in discrete-value scattering matrices of group and fully-connected BD-RIS in SU-MIMO systems was investigated in \cite{nerini2023discretevalue}. Nonetheless, despite these efforts in exploring the impacts of non-idealities for BD-RIS aided systems, architecture discovery for non-ideal BD-RIS remains uninvestigated. To rigorously capture these physical electromagnetic properties and hardware constraints, multiport network theory \cite{pozar2011microwave} has recently emerged as an indispensable modeling framework to characterize mutual coupling and impedance matching in advanced programmable environments, such as RIS \cite{shen2022modeling,gradoni2021endtoend} and Stacked Intelligent Metasurfaces (SIM) \cite{abrardo2025novel}. However, while these existing works primarily utilize multiport networks to model the radiation characteristics of RIS \cite{shen2022modeling,gradoni2021endtoend} or the inter-layer wave propagation of SIM \cite{abrardo2025novel}, the application of such rigorous physical models to discover and optimize the complex interconnections of non-ideal BD-RIS remains an open challenge. More importantly, how non-idealities and circuit complexity jointly affect the performance of BD-RIS remains unclear, making it difficult to achieve the performance-circuit complexity tradeoff in the presence of non-idealities.

Besides architecture discovery for BD-RIS, architecture discovery is also an important research topic in other fields, e.g., protein domain identification and drug design. Architecture discovery in these fields faces profound hurdles due to high-dimensional heterogeneous biological data and an enormous search space created by the combinatorial explosion of possible protein domain arrangements or molecular structures\cite{bernardes2016multiobjective, schneider2020rethinking}. While finding optimal architectures in those disciplines is already inherently difficult, discovering optimal architectures for BD-RIS with non-idealities is substantially more challenging. Beyond the wireless channel dimensions growing linearly and the search space scaling exponentially with $N_I$, evaluating candidate BD-RIS architectures requires solving highly non-convex optimization problems tightly coupled with physical constraints. Consequently, as $N_I$ scales, traditional optimization algorithms become computationally prohibitive and prone to local optima\cite{peng2026lossy}.

Recently, machine learning (ML) has been widely used in wireless communications for channel estimation\cite{zhou2024pay, zhou2025lowoverhead}, beam prediction\cite{li2026outofband}, and beamforming\cite{song2021unsupervised,sobhi-givi2024joint}, etc., by directly learning from complicated wireless channels and maximizing communication-related objectives. However, how to exploit ML to discover optimal architectures for BD-RIS, especially under the severe constraints imposed by hardware non-idealities, remains an uninvestigated open challenge. While traditional optimization algorithms face intractable complexity due to their computationally expensive iterative procedures and the massive search space when jointly optimizing the BD-RIS architecture across a statistical ensemble of channel realizations, ML is able to overcome this computational bottleneck by leveraging neural networks to directly parameterize the highly non-linear mapping between the physical propagation environment and the optimal architecture from this channel ensemble. To fill this research gap, we formulate the learning-based architecture discovery problem and propose a learning-based two-tier architecture discovery framework (LTTADF) for non-ideal BD-RIS architecture discovery, making it possible to achieve the tradeoff between the circuit complexity and the performance of non-ideal BD-RIS. The major contributions of this work can be summarized as follows:
\begin{enumerate}
\item We formulate the learning-based architecture discovery problem to exploit machine learning for discovering optimal architectures for non-ideal BD-RIS with low circuit complexity and, hence, characterize the performance-circuit complexity tradeoff in the presence of non-idealities. We propose the novel LTTADF consisting of an architecture generator and a performance optimizer to jointly learn the probabilities of RIS element interconnections to be beneficial to the performance of BD-RIS, given the circuit complexity $K_{cc}$. By activating the first $K_{cc}$ RIS element interconnections according to the learned probabilities, the optimal BD-RIS architecture with a circuit complexity $K_{cc}$ can be generated by the architecture generator. To avoid getting trapped in poor local optima and thus achieve near-optimal solutions during the performance optimization, we propose a graph-based modeling for BD-RIS and embed the wireless channels into high-dimensional graph representations, where RIS elements are modeled as nodes and RIS element interconnections are modeled as edges. Following this, we propose a residual connection-assisted graph neural network (GNN) to fully exploit the interconnections among RIS elements in high-dimensional representations compliant with both sparse and dense graphs. Different from existing ML-based frameworks for wireless communications, the proposed LTTADF is in a two-tier manner for joint architecture discovery and optimization, which successfully copes with challenges led by the exponentially growing search space and non-convex optimization. It is worth emphasizing that the LTTADF is applicable to both single-user and multi-user cases across all antenna settings (including SISO, MISO, MIMO, etc.), and it can also be extended to discover numerically optimal solutions for other problems that account for non-idealities.

\item We have verified the effectiveness of the proposed LTTADF in cases where optimal architectures have been found via analytical derivations\cite{nerini2024diagonal, zhou2024novel, wu2025beyonddiagonal, nerini2024global}. Our results demonstrate that the performance and circuit complexity of the learned BD-RIS architectures align perfectly with those investigated by analytical derivations in the literature, i.e., learned BD-RIS architectures with the same circuit complexities can achieve nearly the same performance (and numerically even better in some cases) as tree-connected BD-RISs in SU-SISO/SU-MISO systems and band-connected/stem-connected BD-RISs in MU-MIMO systems under ideal BD-RIS considerations, and as tree-connected BD-RISs in SU-SISO systems for BD-RISs with mutual coupling.

\item We have further explored optimal architectures for BD-RIS with non-idealities given specific circuit complexities, specifically BD-RIS with mutual coupling in MU-MIMO systems, lossy BD-RIS in SU-SISO and MU-MIMO systems, and discrete-value BD-RIS in SU-SISO and MU-MIMO systems. Numerical results reveal that mutual coupling at the BD-RIS will not affect the optimal architecture of BD-RIS in MU-MIMO systems. In addition, the results also reveal that increasing circuit complexity can be detrimental for lossy BD-RIS. Therefore, learning the optimal architecture for lossy BD-RIS is meaningful to reduce the circuit complexity and also to avoid negative effects caused by losses at BD-RIS. Moreover, the results based on discrete-value BD-RIS show that circuit complexity can compensate for quantization errors due to fewer quantization bits, and quantization bits can also mitigate performance degradation due to lower circuit complexities.
\end{enumerate}

\textit{Organization}: The remainder of this paper is organized as follows. In Section \ref{Sec. Sys.}, we introduce the system model and formulate the learning-based architecture discovery problem. In Section \ref{Sec. Method}, we propose the LTTADF. In Section \ref{Sec. Sim.}, simulation results are presented to demonstrate the effectiveness of the proposed learning-based architecture discovery framework and reveal optimal architectures for BD-RIS with non-idealities. Finally, we conclude this work in Section \ref{Sec. Con.}.

\textit{Notation}: Bold italic uppercase letter $\boldsymbol{A}$, bold uppercase letter $\mathbf{A}$, and bold lowercase letter $\mathbf{a}$ represent a tensor, a matrix, and a vector, respectively. Calligraphy uppercase letter $\mathcal{A}$ represents a set. $\operatorname{tril}(\mathbf{A})$ and $\operatorname{triu}(\mathbf{A})$ denote the lower and upper triangular elements of $\mathbf{A}$ excluding the diagonal elements. $\mathbf{A}_{m, :}$, $\mathbf{A}_{:,n}$, and $\mathbf{A}_{m,n}$ denote the $m$-th row, the $n$-th column, and the element at the $m$-th row and $n$-th column of $\mathbf{A}$, respectively. $(\cdot)^{T}$, $(\cdot)^{H}$, and $(\cdot)^{-1}$ denote the transpose, conjugate-transpose, and inverse of a matrix, respectively. $\Vert\cdot\Vert_2$ and $\Vert\cdot\Vert_F$ denote the L2 norm and the Frobenius norm, respectively. $\operatorname{vec}(\cdot)$ and $\operatorname{vec}^{-1}(\cdot)$ denote the vectorization operation and its inverse operation, respectively. $\Re\{\cdot\}$ and $\Im\{\cdot\}$ take the real and imaginary parts of the input, respectively. Furthermore, to streamline the presentation across the four considered BD-RIS scenarios, for any generic parameter $\Phi$, we use the base symbol $\Phi$, overline $\overline{\Phi}$, tilde $\widetilde{\Phi}$, and underline $\underline{\Phi}$ to denote its specific instance under the ideal, antenna-coupled (i.e. with mutual coupling between RIS elements), lossy, and discrete-value scenarios, respectively.

\section{System Model and Problem Formulation}\label{Sec. Sys.}

\subsection{System Model}

\begin{figure}[htbp]
\centering
\includegraphics[width=0.7\columnwidth]{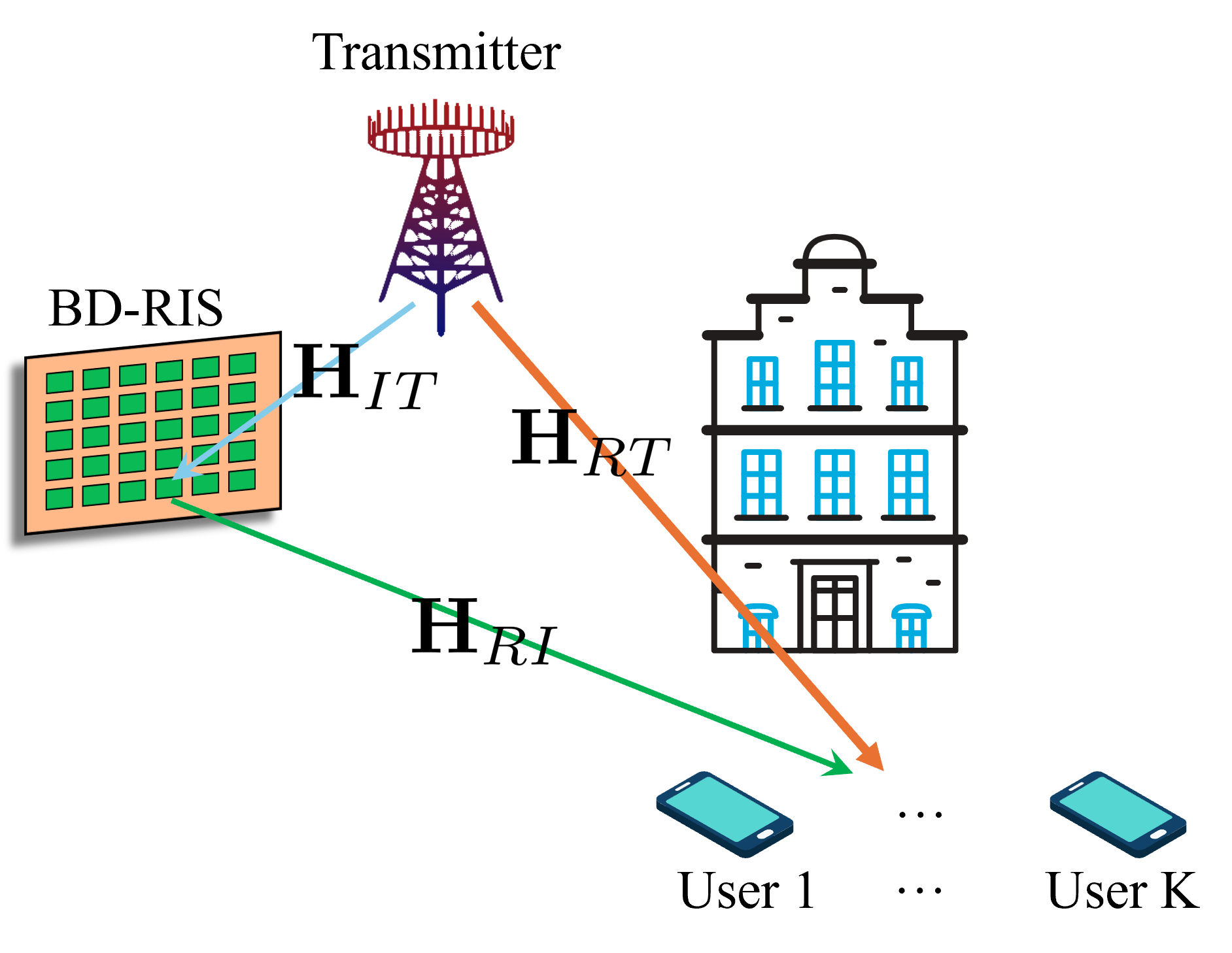}
\caption{The BD-RIS aided MU-MIMO system.}
\label{system}
\end{figure}

As shown in Fig. \ref{system}, we consider a passive BD-RIS aided MU-MIMO system consisting of one transmitter equipped with $N_T$ antennas, one BD-RIS with $N_I$ elements, and $K$ users, where the $k$-th user is equipped with $N_k$ antennas and $N_R = \sum_{k=1}^K N_k$ denotes the total number of user antennas. The $N_I$-element BD-RIS can be modeled as $N_I$ elements connected to an $N_I$-port reconfigurable impedance network consisting of tunable admittance components and characterized by its scattering matrix $\mathbf{\Theta} \in \mathbb{C}^{N_I \times N_I}$. Alternatively, the BD-RIS can also be characterized by the admittance matrix $\mathbf{Y} \in \mathbb{C}^{N_I \times N_I}$ of the reconfigurable impedance network, and $\mathbf{\Theta}$ and $\mathbf{Y}$ are related by\cite{pozar2011microwave}
{\small\begin{equation}
\mathbf{\Theta}=\left(Y_0\mathbf{I}+\mathbf{Y}\right)^{-1}\left(Y_0\mathbf{I}-\mathbf{Y}\right),
\end{equation}
}where $Y_0$ denotes the reference admittance. Generally, we assume the BD-RIS to be reciprocal, which physically implies that the wave transmission characteristics between any two ports are symmetric. Mathematically, this physical property strictly constrains the scattering matrix $\mathbf{\Theta}$ and the admittance matrix $\mathbf{Y}$ to be symmetric (i.e., $\mathbf{\Theta}=\mathbf{\Theta}^T$ and $\mathbf{Y}=\mathbf{Y}^T$).

Under many assumptions of no mutual coupling, perfect matching, unilateral approximation, and no structural scattering \cite{nerini2024universal}, the ideal wireless channel between the transmitter and the users can be expressed as
{\small\begin{equation}\label{channel}
\mathbf{H}  = \mathbf{H}_{RT} + \mathbf{H}_{RI} \mathbf{\Theta} \mathbf{H}_{IT},
\end{equation}
}where $\mathbf{H}_{RT} \in \mathbb{C}^{N_R \times N_T} = [\mathbf{H}_{1,T}^{T}, \ldots, \mathbf{H}_{k,T}^{T}, \ldots, \mathbf{H}_{K,T}^{T}]^{T}$ is the direct channel from the transmitter to the users with $\mathbf{H}_{k,T} \in \mathbb{C}^{N_k \times N_T}$ being the direct channel from the transmitter to the $k$-th user, $\mathbf{H}_{RI} \in \mathbb{C}^{N_R \times N_I} = [\mathbf{H}_{1,I}^{T}, \cdots, \mathbf{H}_{k,I}^{T}, \cdots, \mathbf{H}_{K,I}^{T}]^{T}$ is the channel from the BD-RIS to the users with $\mathbf{H}_{k,I} \in \mathbb{C}^{N_k \times N_I}$ being the channel from the the BD-RIS to the $k$-th user, and $\mathbf{H}_{IT} \in \mathbb{C}^{N_I \times N_T}$ is the channel from the transmitter to the BD-RIS.

To account for mutual coupling at the BD-RIS while retaining the ideal assumptions of perfect matching, unilateral approximation, and no structural scattering \cite{nerini2024universal,nerini2024global}, we need to depart from (\ref{channel}) and introduce the multiport network theory\cite{pozar2011microwave}. Generally, by modeling the wireless channel between the transmitter and the users as an $N_P$-port network, where $N_P=N_T+N_I+N_R$, the wireless channel can be characterized by its admittance matrix $\mathbf{Y}_P \in \mathbb{C}^{N_P \times N_P}$ as
{\small\begin{align}
\mathbf{Y}_P=
\begin{bmatrix}
\mathbf{Y}_{TT} & \mathbf{Y}_{TI} & \mathbf{Y}_{TR}\\
\mathbf{Y}_{IT} & \mathbf{Y}_{II} & \mathbf{Y}_{IR}\\
\mathbf{Y}_{RT} & \mathbf{Y}_{RI} & \mathbf{Y}_{RR}
\end{bmatrix},
\end{align}
}where $\mathbf{Y}_{TT}\in\mathbb{C}^{N_T\times N_T}$, $\mathbf{Y}_{II}\in\mathbb{C}^{N_I\times N_I}$, and $\mathbf{Y}_{RR}\in\mathbb{C}^{N_R\times N_R}$ denote the self-admittance matrices of the antenna arrays at the transmitter, BD-RIS, and users, respectively, $\mathbf{Y}_{RT}\in\mathbb{C}^{N_R\times N_T}$, $\mathbf{Y}_{IT}\in\mathbb{C}^{N_I\times N_T}$, and $\mathbf{Y}_{RI}\in\mathbb{C}^{N_R\times N_I}$ represent the transmission admittance matrices from the transmitter to users, from the transmitter to the BD-RIS, and from the BD-RIS to users, respectively, and $\mathbf{Y}_{TR}\in\mathbb{C}^{N_T\times N_R}$, $\mathbf{Y}_{TI}\in\mathbb{C}^{N_T\times N_I}$, and $\mathbf{Y}_{IR}\in\mathbb{C}^{N_I\times N_R}$ refer to the transmission admittance matrices from the users to the transmitter, from the BD-RIS to the transmitter, and from the users to BD-RIS, respectively. Considering the reciprocity of wireless channels, we have $\mathbf{Y}_{TR}=(\mathbf{Y}_{RT})^T$, $\mathbf{Y}_{TI}=(\mathbf{Y}_{IT})^T$, and $\mathbf{Y}_{IR}=(\mathbf{Y}_{RI})^T$. Denoting $\mathbf{Y}_T \in \mathbb{C}^{N_T \times N_T}$ and $\mathbf{Y}_R \in \mathbb{C}^{N_R \times N_R}$ as the admittance matrices of the $N_T$-port reconfigurable impedance network at the transmitter and the $N_R$-port reconfigurable impedance network at the user side, we assume all the impedances at the transmitter and users to be $Y_0$ and there is no mutual coupling at the transmitter and user sides, implying that $\mathbf{Y}_{T}=Y_0\mathbf{I}$, $\mathbf{Y}_{R}=Y_0\mathbf{I}$, $\mathbf{Y}_{TT}=Y_0\mathbf{I}$, and $\mathbf{Y}_{RR}=Y_0\mathbf{I}$. In addition, we also assume the BD-RIS with mutual coupling to be lossless and reciprocal, indicating that its admittance matrix $\overline{\mathbf{Y}}$ is purely susceptive and symmetric (i.e., $\overline{\mathbf{Y}}=j\overline{\mathbf{B}}$ and $\overline{\mathbf{B}}=\overline{\mathbf{B}}^T$, where $\overline{\mathbf{B}}$ denotes the susceptance matrix). Based on these assumptions, the wireless channel between the transmitter and users, which accounts for mutual coupling at the BD-RIS, can be expressed as\cite{nerini2024universal,nerini2024global}
{\small\begin{equation}\label{mc channel}
\mathbf{H}_{MC} = \overline{\mathbf{S}}_{RT} + \overline{\mathbf{S}}_{RI}\overline{\mathbf{\Theta}}\overline{\mathbf{S}}_{IT},
\end{equation}
}with
{\small\begin{align}
\overline{\mathbf{S}}_{RT} &= -\frac{1}{2Y_0}\left(\mathbf{Y}_{RT}-\frac{\overline{\mathbf{Y}}_{RI}\overline{\mathbf{Y}}_{IT}}{2Y_0}\right),\\
\overline{\mathbf{S}}_{RI} &= -\frac{\overline{\mathbf{Y}}_{RI}}{2Y_0}, \overline{\mathbf{S}}_{IT} = -\frac{\overline{\mathbf{Y}}_{IT}}{2Y_0},\\
\overline{\mathbf{Y}}_{RI} &= \mathbf{Y}_{RI}\Re\{\mathbf{Y}_{II}\}^{-1/2}\sqrt{Y_0},\\
\overline{\mathbf{Y}}_{IT} &= \sqrt{Y_0}\Re\{\mathbf{Y}_{II}\}^{-1/2}\mathbf{Y}_{IT},\\
\overline{\mathbf{\Theta}} &= \left(Y_0\mathbf{I}+j\mathbf{B}^{\prime}\right)^{-1}\left(Y_0\mathbf{I}-j\mathbf{B}^{\prime}\right),\\
\mathbf{B}^{\prime} &= Y_0\Re\{\mathbf{Y}_{II}\}^{-1/2}(\overline{\mathbf{B}}+\Im\{\mathbf{Y}_{II}\})\Re\{\mathbf{Y}_{II}\}^{-1/2},\label{mc B}
\end{align}
}where $\overline{\mathbf{S}}_{RT} \in \mathbb{C}^{N_R \times N_T} = [\overline{\mathbf{S}}_{1,T}^{T}, \ldots, \overline{\mathbf{S}}_{k,T}^{T}, \ldots, \overline{\mathbf{S}}_{K,T}^{T}]^{T}$ is the direct channel from the transmitter to the users with $\overline{\mathbf{S}}_{k,T} \in \mathbb{C}^{N_k \times N_T}$ being the direct channel from the transmitter to the $k$-th user, $\overline{\mathbf{S}}_{RI} \in \mathbb{C}^{N_R \times N_I} = [\overline{\mathbf{S}}_{1,I}^{T}, \cdots, \overline{\mathbf{S}}_{k,I}^{T}, \cdots, \overline{\mathbf{S}}_{K,I}^{T}]^{T}$ is the channel from the BD-RIS to the users with $\overline{\mathbf{S}}_{k,I} \in \mathbb{C}^{N_k \times N_I}$ being the channel from the the BD-RIS to the $k$-th user, and $\overline{\mathbf{S}}_{IT} \in \mathbb{C}^{N_I \times N_T}$ is the channel from the transmitter to the BD-RIS, under mutual coupling assumptions.

Denote $\mathbf{H}_{\text{eff}} = [\mathbf{H}_{\text{eff}, 1}^T,\dots,\mathbf{H}_{\text{eff},k}^T,\dots,\mathbf{H}_{\text{eff},K}^T]^T \in \mathbb{C}^{N_R \times N_T}$ as the effective channel for all $K$ users, where $\mathbf{H}_{\text{eff},k} \in \mathbb{C}^{N_k \times N_T}$ is the effective channel from the transmitter to the $k$-th user following (\ref{channel}) for BD-RIS without mutual coupling (i.e., $\mathbf{H}_{\text{eff}}=\mathbf{H}$) or following (\ref{mc channel}) for BD-RIS with mutual coupling (i.e., $\mathbf{H}_{\text{eff}}=\mathbf{H}_{{MC}}$), respectively. Furthermore, let $s_k \leq \min(N_T, N_k)$ denote the number of downlink streams to the $k$-th user, and practically the total number of streams $N_S = \sum_{k=1}^K s_k \leq N_T$. Denote $\mathbf{P}=[\mathbf{P}_1,\dots,\mathbf{P}_k,\dots,\mathbf{P}_K] \in \mathbb{C}^{N_T\times N_S}$ as the precoding matrix for all $K$ users, where $\mathbf{P}_k\in \mathbb{C}^{N_T\times s_k}$ is the precoding matrix for the $k$-th user. Let $\mathbf{s}_k \in \mathbb{C}^{s_k \times 1}$ represent the transmitted data symbol vector for the $k$-th user with $\mathbb{E}[\mathbf{s}_k \mathbf{s}_k^H] = \mathbf{I}_{s_k}$. The received signal vector $\mathbf{y}_k \in \mathbb{C}^{N_k \times 1}$ at the $k$-th user can be expressed as
{\small\begin{equation}
\mathbf{y}_k = \mathbf{H}_{\text{eff},k} \sum_{k=1}^K \mathbf{P}_k \mathbf{s}_k + \mathbf{n}_k,
\end{equation}
}where $\mathbf{n}_k \sim \mathcal{CN}(\mathbf{0}, \sigma^2 \mathbf{I}_{N_k})$ denotes the additive white Gaussian noise (AWGN) vector at the $k$-th user, with $\sigma^2$ being the noise power.

\subsection{Problem Formulation}\label{Sec. II-B}
Before formulating the BD-RIS architecture discovery problems, we first define the architecture characterization matrix of a BD-RIS.
\begin{definition}[BD-RIS Architecture Characterization Matrix]
The architecture of a BD-RIS with $N_I$ elements can be defined by its architecture characterization matrix $\mathbf{A} \in {\{0,1\}}^{N_I \times N_I}$, where the diagonal elements in $\mathbf{A}$ reflect the connections of the RIS elements to ground via tunable admittance components, and the off-diagonal elements in $\mathbf{A}$ represent the interconnections of RIS elements via tunable admittance components. Specifically, $\mathbf{A}_{i,i}=1$ indicates the $i$-th element is connected to ground, otherwise $\mathbf{A}_{i,i}=0$; $\mathbf{A}_{i,j}=1$ indicates the $i$-th element is connected to the $j$-th element, otherwise $\mathbf{A}_{i,j}=0$. The circuit complexity of the BD-RIS architecture characterized by $\mathbf{A}$ is $\sum_{i=1}^{N_I} \sum_{j=1}^{i} \mathbf{A}_{i,j}$.
\end{definition}

Given the definition of the architecture characterization matrix $\mathbf{A}$, the BD-RIS architecture discovery problem can be described as discovering the most effective BD-RIS architecture characterized by $\mathbf{A}$ and with a circuit complexity of $K_{cc}$ that can achieve the optimal (or numerically near-optimal) performance for a set of $N$ channel realizations.\footnote{The $N$ channel realizations do not represent a temporal sequence of future frames that must be predicted online. Instead, they constitute an offline training ensemble that represents the unbiased statistical distribution of the intended deployment environment.} Note that since the architecture characterization matrix $\mathbf{A}$ is jointly optimized over this entire set of channel realizations rather than on a per-realization basis, the discovered optimal architecture represents a static, hardwired hardware topology designed for a specific propagation environment. Since the topology is fixed at manufacturing, the BD-RIS does not require a dynamic switching network (e.g., PIN diodes or radio-frequency microelectromechanical systems (RF MEMS)) to physically route connections dynamically. Instead, real-time adaptability to a specific channel realization / instantaneous channel state information (CSI) is achieved strictly by adjusting the values of the tunable admittance components.

To establish a universal formulation, we define a generalized performance objective function denoted by $O(\mathbf{H}_\text{eff}, \mathcal{X})$, where $\mathcal{X}$ denotes the set of objective-specific auxiliary variables subject to optional objective-specific constraints $C(\mathcal{X}) \leq 0$. Depending on the specific system setup, $O(\cdot)$ can be instantiated as various performance metrics. For example, in single-user cases, $O(\cdot)$ can be instantiated as the channel gain as
{\small\begin{equation}
G(\mathbf{H}_\text{eff}) = \Vert \mathbf{H}_\text{eff} \Vert ^{2}_2,
\end{equation}
}with $\mathcal{X} = \emptyset$. Alternatively, in multi-user cases, $O(\cdot)$ can be instantiated as the sum rate of all $K$ users given by
{\small\begin{equation}
R_{\text{sum}}(\mathbf{P}, \mathbf{H}_\text{eff}) = \sum_{k=1}^K R_k(\mathbf{P},\mathbf{H}_{\text{eff},k}),
\end{equation}
}where $R_k(\mathbf{P},\mathbf{H}_{\text{eff},k})$ calculates the achievable rate of the $k$-th user as
{\small\begin{align}
R_k(\mathbf{P}, \mathbf{H}_{\text{eff},k})& =\log_2\det \biggl(\mathbf{I} + \mathbf{H}_{\text{eff},k} \mathbf{P}_k \mathbf{P}_k^H \mathbf{H}_{\text{eff},k}^H \nonumber \\
\quad \times &\bigl(\sum_{j \neq k} \mathbf{H}_{\text{eff},k} \mathbf{P}_j \mathbf{P}_j^H \mathbf{H}_{\text{eff},k}^H + \sigma^2 \mathbf{I}\bigr)^{-1}\biggr),
\end{align}
}and $\mathcal{X} = \{\mathbf{P}\}$ subject to the transmit power constraint $\Vert \mathbf{P} \Vert_F^2 - P_T \leq 0$ with $P_T$ being the maximum transmit power at the transmitter. Consequently, the architecture discovery problems under the ideal, antenna-coupled, lossy, and discrete-value BD-RIS scenarios can be formulated as follows. For clarity of presentation, we establish a unified convention to distinguish the four considered BD-RIS scenarios. For any general variable, set, or function (denoted generically as $\Phi$), its specific representations under the ideal, antenna-coupled, lossy, and discrete-value scenarios are strictly denoted by $\Phi$, $\overline{\Phi}$, $\widetilde{\Phi}$, and $\underline{\Phi}$, respectively. For instance, the generalized performance objective $O(\cdot)$ and the set of auxiliary variables $\mathcal{X}$ will naturally follow this convention (e.g., $O(\cdot)$ and $\mathcal{X}$ for the ideal case, $\overline{O}(\cdot)$ and $\overline{\mathcal{X}}$ for the mutual coupling case, etc.).

\subsubsection{Architecture Discovery for Ideal BD-RIS}
Considering lossless and reciprocal BD-RIS with purely susceptive $\mathbf{Y}$ (i.e., $\mathbf{Y}=j\mathbf{B}$ and $\mathbf{B}=\mathbf{B}^T$, where $\mathbf{B}$ denotes the susceptance matrix), discovering a BD-RIS architecture characterized by $\mathbf{A}$ and with a circuit complexity of $K_{cc}$ that maximizes the averaged performance objective for all $N$ channel realizations $\{ \mathbf{H}_{RT}^{(n)}, \mathbf{H}_{RI}^{(n)}, \mathbf{H}_{IT}^{(n)} \}_{n=1}^{N}$ can be formulated as
{\small\begin{subequations}
\begin{align}
& \max_{\mathbf{A},{\{\mathbf{B}^{(n)}\}}_{n=1}^N, \mathcal{X}} \frac{1}{N} \sum_{n=1}^{N} O(\mathbf{H}_\text{eff}^{(n)}, \mathcal{X}^{(n)}), \\
\mathsf{\mathrm{s.t.}} \ \ \ 
& \sum_{i=1}^{N_I} \sum_{j=1}^{i} \mathbf{A}_{i,j} = K_{cc},\\
& \mathbf{B}^{(n)}_{i,j}\begin{cases}
    \neq 0, & \mathbf{A}_{i,j} = 1, \\
    =0,  & \mathbf{A}_{i,j} = 0,
\end{cases} \label{ideal_const_B_A} \\
& \mathbf{B}^{(n)} = ({\mathbf{B}^{(n)}})^T, \\
& \mathbf{Y}^{(n)}=j\mathbf{B}^{(n)},  \\
& \mathbf{\Theta}^{(n)}=\left(Y_0\mathbf{I}+\mathbf{Y}^{(n)}\right)^{-1}\left(Y_0\mathbf{I}-\mathbf{Y}^{(n)}\right), \\
& \mathbf{H}_\text{eff}^{(n)}  = \mathbf{H}_{RT}^{(n)} + \mathbf{H}_{RI}^{(n)} \mathbf{\Theta}^{(n)} \mathbf{H}_{IT}^{(n)},\\
& C(\mathcal{X}^{(n)}) \leq 0. \label{ideal_const_X}
\end{align}
\end{subequations}}

\subsubsection{Architecture Discovery for BD-RIS with Mutual Coupling}

According to (\ref{mc channel}) - (\ref{mc B}), the architecture discovery problem for lossless and reciprocal BD-RIS with mutual coupling (i.e., $\overline{\mathbf{Y}}=j\overline{\mathbf{B}}$ and $\overline{\mathbf{B}}=\overline{\mathbf{B}}^T$) given $N$ channel realizations $\{ \overline{\mathbf{S}}_{RT}^{(n)}, \overline{\mathbf{S}}_{RI}^{(n)}, \overline{\mathbf{S}}_{IT}^{(n)} \}_{n=1}^{N}$ can be formulated as
{\small\begin{subequations}
\begin{align}
& \max_{\overline{\mathbf{A}},{\{\overline{\mathbf{B}}^{(n)}\}}_{n=1}^N, \overline{\mathcal{X}}} \frac{1}{N} \sum_{n=1}^{N} \overline{O}(\overline{\mathbf{H}}_\text{eff}^{(n)}, \overline{\mathcal{X}}^{(n)}), \\
\mathsf{\mathrm{s.t.}} \ \ \ 
& \sum_{i=1}^{N_I} \sum_{j=1}^{i} \overline{\mathbf{A}}_{i,j} = K_{cc},\\
& \overline{\mathbf{B}}^{(n)}_{i,j}\begin{cases}
    \neq 0, & \overline{\mathbf{A}}_{i,j} = 1,\\
    =0,  & \overline{\mathbf{A}}_{i,j} = 0,
\end{cases}\\
& \overline{\mathbf{B}}^{(n)} = ({\overline{\mathbf{B}}^{(n)}})^T,\\
&\mathbf{B}^{\prime(n)} = Y_0\Re\{\mathbf{Y}_{II}^{(n)}\}^{-1/2}(\overline{\mathbf{B}}^{(n)}+\Im\{\mathbf{Y}_{II}^{(n)}\})\Re\{\mathbf{Y}_{II}^{(n)}\}^{-1/2},\\
&\overline{\mathbf{\Theta}}^{(n)}=\left(Y_0\mathbf{I}+j\mathbf{B}^{\prime(n)}\right)^{-1}\left(Y_0\mathbf{I}-j\mathbf{B}^{\prime(n)}\right),\\
& \overline{\mathbf{H}}_\text{eff}^{(n)}  = \overline{\mathbf{S}}_{RT}^{(n)} + \overline{\mathbf{S}}_{RI}^{(n)} \overline{\mathbf{\Theta}}^{(n)} \overline{\mathbf{S}}_{IT}^{(n)},\\
& C(\overline{\mathcal{X}}^{(n)}) \leq 0.
\end{align}
\end{subequations}}

\subsubsection{Architecture Discovery for Lossy BD-RIS}

Lossy BD-RIS is modeled in \cite{peng2026lossy} as
{\small\begin{equation}
\widetilde{\mathbf{Y}}_{i,j} = \begin{cases}
-\widetilde{Y}_{ij}, &i\ne j, \\
\sum_{k=1}^{N_I}\widetilde{Y}_{ik}, & i = j,
\end{cases}
\end{equation}}
where
{\small\begin{align}
& \Re\{\widetilde{Y}_{ij}\} = \frac{R}{R^2 + \left(\omega L_2 - \frac{1}{\omega \widetilde{C}_{ij}} \right)^2},\\
& \Im\{\widetilde{Y}_{ij}\} = \left( -\frac{1}{\omega L_1} + \frac{-\omega L_2 + \frac{1}{\omega \widetilde{C}_{ij}}}{R^2 + \left(\omega L_2 - \frac{1}{\omega \widetilde{C}_{ij}} \right)^2} \right),
\end{align}
}where $\widetilde{C}_{ij}$ is the capacitance associated with the tunable admittance $\widetilde{Y}_{ij}$, $L_1$ and $L_2$ are two inductances, and $R$ is a resistor which characterizes the loss of lossy BD-RIS. For clarity, we define the capacitance matrix $\widetilde{\mathbf{C}} \in \mathbb{R}^{N_I \times N_I}$ with $\widetilde{\mathbf{C}}_{i,j}=\widetilde{C}_{ij}$.
By adopting this lossy BD-RIS modeling,  the architecture discovery problem for lossy and reciprocal BD-RIS (i.e., $\widetilde{\mathbf{Y}} = ({\widetilde{\mathbf{Y}}})^T)$ given $N$ channel realizations $\{ \mathbf{H}_{RT}^{(n)}, \mathbf{H}_{RI}^{(n)}, \mathbf{H}_{IT}^{(n)} \}_{n=1}^{N}$ can be formulated as
{\small\begin{subequations}
\begin{align}
& \max_{\widetilde{\mathbf{A}},{\{\widetilde{\mathbf{C}}^{(n)}\}}_{n=1}^N,\widetilde{\mathcal{X}}} \frac{1}{N} \sum_{n=1}^{N} \widetilde{O}(\widetilde{\mathbf{H}}_\text{eff}^{(n)}, \widetilde{\mathcal{X}}^{(n)}), \label{lossy PF}\\
\mathsf{\mathrm{s.t.}} \ \ \
& \sum_{i=1}^{N_I} \sum_{j=1}^{i} \widetilde{\mathbf{A}}_{i,j} = K_{cc},\\
& \widetilde{\mathbf{C}}^{(n)}_{i,j}\begin{cases}
    \neq 0, & \widetilde{\mathbf{A}}_{i,j} = 1,\\
    =0,  & \widetilde{\mathbf{A}}_{i,j} = 0,
\end{cases}\\
& \Re\{\widetilde{Y}^{(n)}_{ij}\} = \frac{R}{R^2 + \left(\omega L_2 - \frac{1}{\omega \widetilde{\mathbf{C}}^{(n)}_{i,j}} \right)^2},\\
& \Im\{\widetilde{Y}^{(n)}_{ij}\} = \left( -\frac{1}{\omega L_1} + \frac{-\omega L_2 + \frac{1}{\omega \widetilde{\mathbf{C}}^{(n)}_{i,j}}}{R^2 + \left(\omega L_2 - \frac{1}{\omega \widetilde{\mathbf{C}}^{(n)}_{i,j}} \right)^2} \right),\\
& \widetilde{\mathbf{Y}}^{(n)}_{i,j} = \begin{cases}
-\widetilde{Y}^{(n)}_{ij}, &\widetilde{\mathbf{A}}_{i,j} = 1 \text{ and } i\ne j , \\
\sum_{k=1}^{N_I}\widetilde{Y}^{(n)}_{ik}, &\widetilde{\mathbf{A}}_{i,j} = 1 \text{ and }  i = j, \\
0,  & \widetilde{\mathbf{A}}_{i,j} = 0,
\end{cases} \\
& \widetilde{\mathbf{Y}}^{(n)} = ({\widetilde{\mathbf{Y}}^{(n)}})^T,\\
& \widetilde{\mathbf{\Theta}}^{(n)}=\left(Y_0\mathbf{I}+\widetilde{\mathbf{Y}}^{(n)}\right)^{-1}\left(Y_0\mathbf{I}-\widetilde{\mathbf{Y}}^{(n)}\right), \\
& \widetilde{\mathbf{H}}_\text{eff}^{(n)}  = \mathbf{H}_{RT}^{(n)} + \mathbf{H}_{RI}^{(n)} \widetilde{\mathbf{\Theta}}^{(n)} \mathbf{H}_{IT}^{(n)}, \\
& C(\widetilde{\mathcal{X}}^{(n)}) \leq 0.
\end{align}
\end{subequations}}

\subsubsection{Architecture Discovery for Discrete-Value BD-RIS}

We then consider the architecture discovery for lossless and reciprocal discrete-value BD-RIS with $\underline{\mathbf{Y}}=j\underline{\mathbf{B}}$ and $\underline{\mathbf{B}} = \underline{\mathbf{B}}^T$. Different from continuous BD-RIS whose susceptance matrix entries are allowed to be arbitrary real values, the susceptance matrix entries of lossless discrete-value BD-RIS can only be selected from an $N_b$-bit codebook $\mathcal{B}^{N_b}$ with $2^{N_b}$ codewords. Therefore, the architecture discovery problem for discrete-value BD-RIS quantized by $\mathcal{B}^{N_b}$ given $N$ channel realizations $\{ \mathbf{H}_{RT}^{(n)}, \mathbf{H}_{RI}^{(n)}, \mathbf{H}_{IT}^{(n)} \}_{n=1}^{N}$ can be formulated as
{\small\begin{subequations}
\begin{align}
&\max_{\underline{\mathbf{A}}, \mathcal{B}^{N_b},{\{\underline{\mathbf{B}}^{(n)}\}}_{n=1}^N,\underline{\mathcal{X}}} \frac{1}{N} \sum_{n=1}^{N} \underline{O}(\underline{\mathbf{H}}_\text{eff}^{(n)}, \underline{\mathcal{X}}^{(n)}), \label{discrete PF}\\
\mathsf{\mathrm{s.t.}} \ \ \ 
& \sum_{i=1}^{N} \sum_{j=1}^{i} \underline{\mathbf{A}}_{i,j} = K_{cc}, \\
& \underline{\mathbf{B}}^{(n)}_{i,j} \in \begin{cases}
    \mathcal{B}^{N_b}, & \underline{\mathbf{A}}_{i,j} = 1, \\
    \{0\},  & \underline{\mathbf{A}}_{i,j} = 0,
\end{cases} \\
& \underline{\mathbf{B}}^{(n)} = (\underline{\mathbf{B}}^{(n)})^T, \\
& \underline{\mathbf{\Theta}}^{(n)}=\left(Y_0\mathbf{I}+j\underline{\mathbf{B}}^{(n)}\right)^{-1}\left(Y_0\mathbf{I}-j\underline{\mathbf{B}}^{(n)}\right),\\
& \underline{\mathbf{H}}_\text{eff}^{(n)}  = \mathbf{H}_{RT}^{(n)} + \mathbf{H}_{RI}^{(n)} \underline{\mathbf{\Theta}}^{(n)} \mathbf{H}_{IT}^{(n)},\\
& C(\underline{\mathcal{X}}^{(n)}) \leq 0.
\end{align}
\end{subequations}}

\section{Learning-Based BD-RIS Architecture Discovery}\label{Sec. Method}
\begin{figure*}[htbp]
\centering
\includegraphics[width=\textwidth]{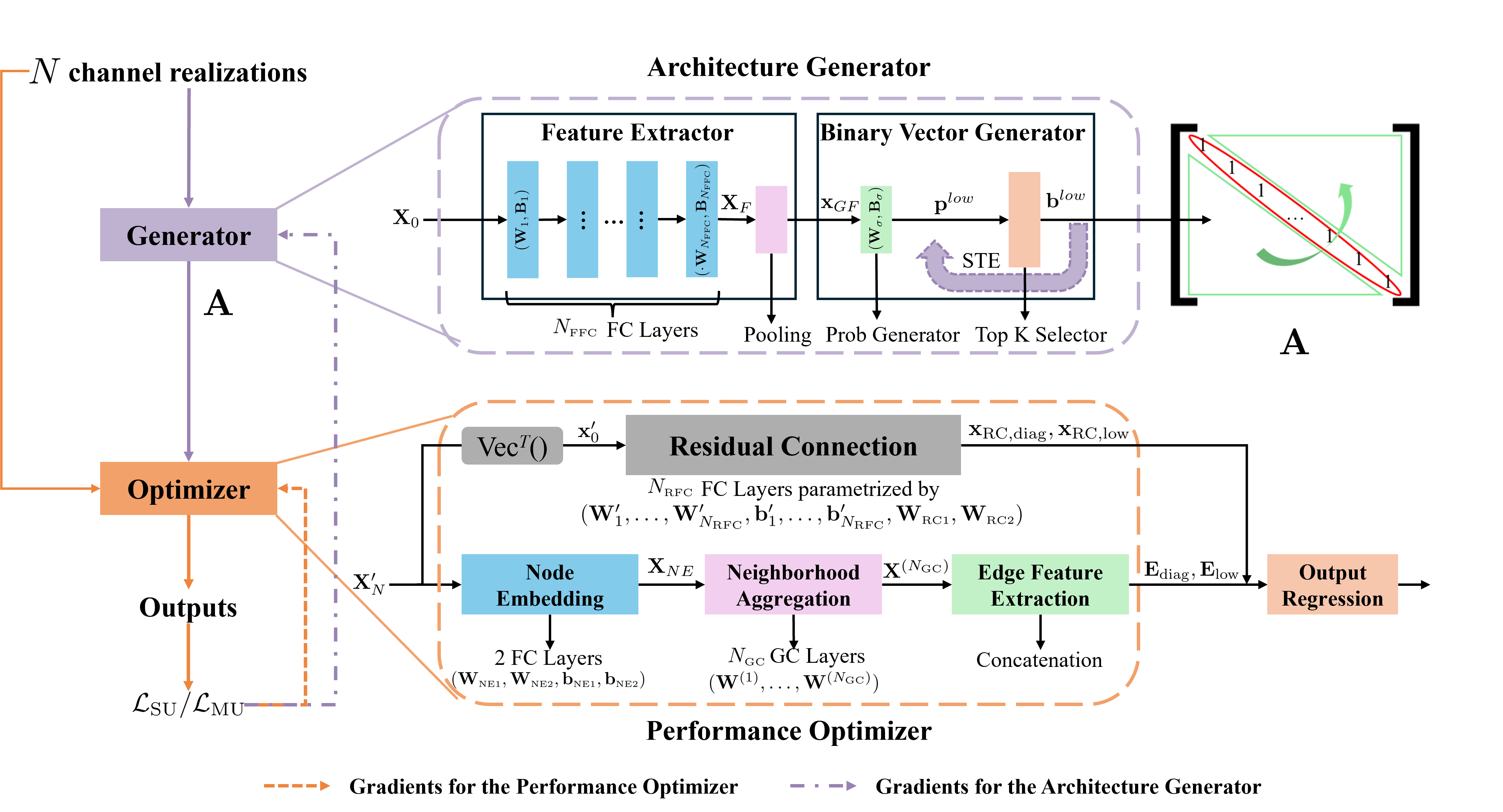}
\caption{The learning-based two-tier architecture discovery framework (LTTADF).}
\label{framework}
\end{figure*}

To address the challenges arising from intractable global architecture search and difficulties in analytically or numerically achieving optimal solutions via traditional mathematical tools, we resort to machine learning to effectively explore all possible BD-RIS architectures and determine the optimal BD-RIS architecture with a certain circuit complexity. Specifically, we propose the learning-based two-tier architecture discovery framework shown in Fig. \ref{framework}, which consists of an architecture generator and a performance optimizer.\footnote{Note that unless otherwise specified, in Section \ref{Sec. Method} and Fig. \ref{framework}, we use the notations for the ideal case to ease the presentation.} The architecture generator generates a BD-RIS architecture, and then the performance optimizer evaluates the performance of the generated BD-RIS architecture and feeds this information back to the architecture generator. By doing so, the architecture generator and performance optimizer jointly learn the probabilities of RIS element interconnections to be beneficial to the performance of the BD-RIS, such that the optimal BD-RIS architecture with a certain circuit complexity can be determined.

\subsection{Architecture Generator}
\subsubsection{Feature Extraction}
The architecture generator aims to generate a BD-RIS architecture characterized by the architecture characterization matrix $\mathbf{A}$ from $N$ input channel realizations. To this end, $N_\text{\tiny{FFC}}$ fully-connected (FC) layers are first designed to extract high-dimensional features from the inputs, whose output $\mathbf{X}_{F} \in \mathbb{R}^{N \times d_{N_\text{\tiny{FFC}}}}$ can thus be expressed as
\begin{equation}
\mathbf{X}_{F} = \operatorname{ReLU}(\operatorname{ReLU}(\mathbf{X}_{0}\mathbf{W}_{1}+\mathbf{B}_{1})\cdots\mathbf{W}_{N_\text{\tiny{FFC}}}+\mathbf{B}_{N_\text{\tiny{FFC}}}),
\end{equation}
where $\operatorname{ReLU}(\cdot)$ denotes the rectified linear unit activation function, $\mathbf{W}_{1} \in \mathbb{R}^{2N_I(N_T+N_R) \times d_1}$, $\ldots$, $\mathbf{W}_{N_\text{\tiny{FFC}}} \in \mathbb{R}^{d_{N_\text{\tiny{FFC}} - 1} \times d_{N_\text{\tiny{FFC}}}}$ and $\mathbf{B}_{1} \in \mathbb{R}^{N \times d_1}$, $\ldots$, $\mathbf{B}_{N_\text{\tiny{FFC}}} \in \mathbb{R}^{N \times d_{N_\text{\tiny{FFC}}}}$ represent learnable weight and bias matrices corresponding to the $N_\text{\tiny{FFC}}$ FC layers respectively, and $\mathbf{X}_{0} = [\mathbf{x}^{(1)}_0, \ldots, \mathbf{x}^{(n)}_0, \ldots, \mathbf{x}^{(N)}_0]^T \in \mathbb{R}^{N \times 2N_I(N_T+N_R)}$ is the input matrix from $N$ channel realizations. Note that for ideal, lossy, and discrete-value BD-RISs, the channel model in (\ref{channel}) is considered, indicating that the BD-RIS is involved in the transmitter-RIS channel $\mathbf{H}_{IT}$ and the RIS-user channel $\mathbf{H}_{RI}$. Therefore, $\mathbf{x}^{(n)}_0=[\operatorname{vec}(\Re(\mathbf{H}_{IT}^{(n)})),\operatorname{vec}(\Im(\mathbf{H}_{IT}^{(n)})),\operatorname{vec}(\Re(\mathbf{H}_{RI}^{(n)})),\operatorname{vec}(\Im(\mathbf{H}_{RI}^{(n)}))] \in \mathbb{R}^{ 2N_I(N_T+N_R) \times 1}$ where $\operatorname{vec}(\cdot)$ denotes the vectorization operation. While for BD-RIS with mutual coupling, the channel model accounting for mutual coupling in (\ref{mc channel}) is considered, indicating that the BD-RIS is involved in the transformed channels $\overline{\mathbf{S}}_{IT}$ and $\overline{\mathbf{S}}_{RI}$. Therefore, $\mathbf{x}^{(n)}_0 = [\operatorname{vec}(\Re(\overline{\mathbf{S}}_{IT}^{(n)})), \operatorname{vec}(\Im(\overline{\mathbf{S}}_{IT}^{(n)})), \operatorname{vec}(\Re(\overline{\mathbf{S}}_{RI}^{(n)})), \operatorname{vec}(\Im(\overline{\mathbf{S}}_{RI}^{(n)}))] \in \mathbb{R}^{ 2N_I(N_T+N_R)\times 1}$.

Then an average pooling layer is designed to extract a realization-wise global feature $\mathbf{x}_{GF} \in \mathbb{R}^{1 \times d_{N_\text{\tiny{FFC}}}}$ as
\begin{equation}
\mathbf{x}_{GF} = \frac{1}{N} \sum_{i=1}^{N} (\mathbf{X}_{F})_{i,:}.
\end{equation}

\subsubsection{BD-RIS Architecture Generation}
The realization-wise global feature $\mathbf{x}_{GF}$ can then be used to generate a BD-RIS architecture. Since only reciprocal BD-RIS is considered in this paper, the architecture characterization matrix $\mathbf{A}$ is split into three parts: the diagonal elements, the lower off-diagonal elements, and the upper off-diagonal elements, where the upper off-diagonal elements are the transpose of the lower off-diagonal elements. In addition, we practically consider single-connected BD-RIS as the BD-RIS with the lowest circuit complexity that we are interested in, which indicates that the diagonal elements of $\mathbf{A}$ can be set as $1$ across the learning process. Therefore, the realization-wise global feature $\mathbf{x}_{GF}$ is used to generate only a binary vector representing the lower off-diagonal elements of $\mathbf{A}$. After that, the lower off-diagonal elements of $\mathbf{A}$ are transposed to form the upper off-diagonal elements of $\mathbf{A}$, such that the reciprocal constraint is satisfied.

Since binary vectors are discrete and non-differentiable, it is quite challenging to directly learn the desired binary vector using neural networks. To solve this issue, we propose to first learn a probability vector indicating the probabilities of each lower off-diagonal element to be beneficial to the performance of the BD-RIS. Then, based on the learned probabilities, a BD-RIS architecture that is able to achieve the highest performance and satisfy the circuit complexity $K_{cc}$ can be determined by activating the top $K_{cc}$ lower off-diagonal elements with the highest probabilities. Nonetheless, it is worth noting that the top-$K_{cc}$ selection is also non-differentiable, which requires further design for the architecture generator to learn from the channel realizations well. Therefore, we propose the binary vector generator by combining a probability generator, a top-$K_{cc}$ selector $\operatorname{TopK}(\cdot, \cdot)$, and a straight-through estimator (STE)\cite{bengio2013estimating} dedicated to solving non-differentiable issues. The probability generator is comprised of a linear layer and a Sigmoid function denoted by $\sigma(\cdot)$, whose output is
\begin{align}
\mathbf{p}^{low} = \sigma(\mathbf{x}_{GF}\mathbf{W}_{\sigma}+\mathbf{b}_{\sigma}),
\end{align}
where $\mathbf{p}^{low} \in \mathbb{R}^{1 \times N_I(N_I - 1)/2}$ is the probability vector, $\mathbf{W}_{\sigma} \in \mathbb{R}^{d_{N_\text{\tiny{FFC}}} \times N_I(N_I - 1)/2}$ and $\mathbf{b}_{\sigma} \in \mathbb{R}^{1 \times N_I(N_I - 1)/2}$ are learnable weight matrix and bias vector corresponding to the probability generator.
The top-$K_{cc}$ selector $\operatorname{TopK}(\mathbf{p}^{low}, K_{cc})$ then determines $K_{cc}$ lower off-diagonal elements to be activated based on the generated probability vector $\mathbf{p}^{low}$, forming the binary vector $\mathbf{b}^{low} \in {\{0,1\}}^{1 \times N_I(N_I - 1)/2}$ whose $K_{cc}$ elements are $1$s and the other $N_I(N_I - 1)/2 - K_{cc}$ elements are $0$s as
\begin{equation}
\mathbf{b}^{low}_i =
\begin{cases}
1, & \mathbf{p}^{low}_i \in \operatorname{TopK}(\mathbf{p}^{low}, K_{cc}), \\
0, & \text{otherwise}.
\end{cases}
\end{equation}
Then, to form $\mathbf{A}$, we further set
\begin{align}
\mathbf{A} &= \mathbf{I}_{N_I},\\
\operatorname{tril}(\mathbf{A}) &= \mathbf{b}^{low},\\
\operatorname{triu}(\mathbf{A}) &= (\operatorname{tril}(\mathbf{A}))^T,
\end{align}
where $\operatorname{tril}(\mathbf{A})$ and $\operatorname{triu}(\mathbf{A})$ denote the lower and upper triangular elements of $\mathbf{A}$ excluding the diagonal elements, respectively.
Finally, the STE is exploited to provide the gradient for learning from the channel realizations and updating the architecture generator, which approximates the gradient as $\frac{\partial \mathbf{b}^{low}}{\partial \mathbf{p}^{low}} \approx 1$.

\subsection{Performance Optimizer}
Given the generated BD-RIS architecture characterized by $\mathbf{A}$, the performance optimizer aims to maximize the objectives, such that the performance of the generated BD-RIS architecture is evaluated and fed back to the architecture generator for further learning. Since ideal BD-RIS has primarily been modeled as a graph to enable architecture discovery based on graph theory \cite{nerini2024diagonal,wu2025beyonddiagonal}, we also model the BD-RIS as a graph following the definition of ``Graph''\cite{ma2021deep}.

\begin{definition}[Graph]
A graph is defined as $\mathcal{G}=\{\mathcal{V}, \mathcal{E}\}$, where $\mathcal{V}=\{v_1,\ldots,v_{N_v}\}$ denotes a set of $N_v$ nodes and $\mathcal{E}=\{e_1,\ldots,e_{N_e}\}$ denotes a set of $N_e$ edges.
\end{definition}

According to this definition, an $N_I$-element BD-RIS can be modeled as a graph with $N_I$ nodes and $N_E$ edges, where each element is regarded as a node and each tunable admittance component connecting two elements is regarded as an edge.

Since maximizing the objectives for BD-RIS with non-idealities is challenging, we design a GNN-based performance optimizer to fully exploit the interconnections among RIS elements in high-dimensional representations, aiming to avoid getting trapped in poor local optima\cite{amos2017optnet, li2016learning}. Note that despite the graph modeling in \cite{nerini2024diagonal,wu2025beyonddiagonal}, a graph representation to enable the exploitation of GNNs remains unexplored. Therefore, in the following parts, we first propose a graph representation for BD-RIS, and then the learning-based two-tier architecture discovery framework is proposed based on such a graph representation.

\subsubsection{Graph Representation}
The graph representation for GNNs includes an adjacency matrix reflecting the graph's topology and a node feature matrix reflecting all nodes' features. The adjacency matrix of a graph can be defined as follows \cite{ma2021deep}.

\begin{definition}[Adjacency Matrix]
For a given graph $\mathcal{G}=\{\mathcal{V}, \mathcal{E}\}$ with $N_v$ nodes, its adjacency matrix is defined as $\mathbf{A}_{\mathcal{G}} \in {\{0,1\}}^{N_v \times N_v}$, where $(\mathbf{A}_{\mathcal{G}})_{i,j}$ represents the connectivity between the two nodes $v_i$ and $v_j$. Specifically, $(\mathbf{A}_{\mathcal{G}})_{i,j}=1$ if the node $v_i$ is connected to $v_j$, otherwise $(\mathbf{A}_{\mathcal{G}})_{i,j}=0$.
\end{definition}

It is worth emphasizing that the architecture characterization matrix $\mathbf{A}$ of a BD-RIS matches well with the adjacency matrix $\mathbf{A}_{\mathcal{G}}$ of the graph associated with the BD-RIS, with $\mathbf{A}_{i,i}=1$ indicates a connection to ground whereas $(\mathbf{A}_{\mathcal{G}})_{i,i}=1$ indicates a self-loop in the node $v_i$, and $\mathbf{A}_{i,j}=1$ indicates a connection between the $i$-th element and the $j$-th element via a tunable admittance component whereas $(\mathbf{A}_{\mathcal{G}})_{i,j}=1$ indicates a connection between $v_i$ and $v_j$. As such, the architecture characterization matrix $\mathbf{A}$ generated by the architecture generator can be directly used as the adjacency matrix of the BD-RIS graph, i.e., $\mathbf{A}_{\mathcal{G}}=\mathbf{A}$.

As for the node feature matrix, two different cases should be considered:
\begin{itemize}
\item Ideal, lossy, and discrete-value BD-RISs: The BD-RIS is involved in the transmitter-RIS channel $\mathbf{H}_{IT}$ and the RIS-user channel $\mathbf{H}_{RI}$ according to (\ref{channel}). Therefore, the node feature matrix $\mathbf{X}_N$ can be obtained as
\begin{align}
\mathbf{X}_{IT} &\in \mathbb{R}^{N_I \times 2N_T} = \left[\Re(\mathbf{H}_{IT}), \Im(\mathbf{H}_{IT})\right],\\
\mathbf{X}_{RI} &\in \mathbb{R}^{N_I \times 2N_R} = \left[\Re(\mathbf{H}_{RI}^T), \Im(\mathbf{H}_{RI}^T)\right],\\
\mathbf{X}_N &\in \mathbb{R}^{N_I \times 2(N_T+N_R)} = \left[ \mathbf{X}_{IT}, \mathbf{X}_{RI}\right].
\end{align}

\item BD-RIS with mutual coupling: The BD-RIS is involved in the transformed channels $\overline{\mathbf{S}}_{IT}$ and $\overline{\mathbf{S}}_{RI}$ according to (\ref{mc channel}). Therefore, the node feature matrix $\overline{\mathbf{X}}_N$ can be obtained as
\begin{align}
\overline{\mathbf{X}}_{IT} &\in \mathbb{R}^{N_I \times 2N_T} = \left[\Re(\overline{\mathbf{S}}_{IT}), \Im(\overline{\mathbf{S}}_{IT})\right],\\
\overline{\mathbf{X}}_{RI} & \in \mathbb{R}^{N_I \times 2N_R} = \left[\Re(\overline{\mathbf{S}}_{RI}^T), \Im(\overline{\mathbf{S}}_{RI}^T)\right],\\
\overline{\mathbf{X}}_N &\in \mathbb{R}^{N_I \times 2(N_T+N_R)} = \left[ \overline{\mathbf{X}}_{IT}, \overline{\mathbf{X}}_{RI}\right].
\end{align}
\end{itemize}

We use two node embedding layers to further extract high-dimensional node embeddings $\mathbf{X}_{NE} \in \mathbb{R}^{N_I \times d_\text{\tiny{NE2}}}$ given $\mathbf{X}_{N}^\prime=\mathbf{X}_{N}$ or $\mathbf{X}_{N}^\prime=\overline{\mathbf{X}}_{N} $ as
\begin{equation}
\mathbf{X}_{NE} = \operatorname{ReLU}(\operatorname{ReLU}(\mathbf{X}_{N}^\prime\mathbf{W}_\text{\tiny{NE1}}+\mathbf{b}_\text{\tiny{NE1}})\mathbf{W}_\text{\tiny{NE2}}+\mathbf{b}_\text{\tiny{NE2}}),
\end{equation}
where $\mathbf{W}_\text{\tiny{NE1}} \in \mathbb{R}^{2(N_T+N_R) \times d_\text{\tiny{NE1}}}$ and $\mathbf{W}_\text{\tiny{NE2}} \in \mathbb{R}^{d_\text{\tiny{NE1}} \times d_\text{\tiny{NE2}}}$ are learnable weight matrices, and $\mathbf{b}_\text{\tiny{NE1}} \in \mathbb{R}^{1 \times d_\text{\tiny{NE1}}}$ and $\mathbf{b}_\text{\tiny{NE2}} \in \mathbb{R}^{1 \times d_\text{\tiny{NE2}}}$ are learnable bias vectors.
 
\subsubsection{Neighborhood Aggregation and Edge Feature Extraction}

According to the graph modeling for BD-RIS, each tunable admittance component connecting two elements is regarded as an edge. Therefore, the BD-RIS performance optimization is directly connected to edge-level features of the graph. The major benefit of GNNs lies in their capability to obtain node-level features via neighborhood aggregation according to the graph's topology, such that edge-level features can then be obtained accordingly from node-level features. To this end, we design specific GNN-based modules to obtain node-level features and edge-level features as follows.

To capitalize on neighborhood aggregation, we design a neighborhood aggregation module consisting of $N_\text{\tiny{GC}}$ graph convolutional (GC) layers\cite{kipf2016semi}, where the output of the $l$-th layer, $l = 1, 2, \ldots, N_\text{\tiny{GC}}$, denoted by $\mathbf{X}^{(l)} \in \mathbb{R}^{N_I \times d_\text{\tiny{GC,l}}}$, can be expressed as
\begin{align}
\mathbf{X}^{(l)} = \operatorname{ReLU}(\mathbf{D}^{-1 / 2} \mathbf{A}_{\mathcal{G}} \mathbf{D}^{-1 / 2} \mathbf{X}^{(l-1)} \mathbf{W}^{(l-1)}),
\end{align}
where $\mathbf{X}^{(0)}$ = $\mathbf{X}_{NE}$, $\mathbf{W}^{(l)} \in \mathbb{R}^{d_\text{\tiny{GC,l-1}} \times d_\text{\tiny{GC,l}}}$ is the learnable weight matrix of the $l$-th layer, and $\mathbf{D}$ is the degree matrix determined by
\begin{equation}
\mathbf{D}_{i,j}=\left\{\begin{array}{cc}
\sum_{j=1}^{N_I} (\mathbf{A}_{\mathcal{G}})_{i,j}, & i=j, \\
0, & \text { otherwise. }
\end{array}\right.
\end{equation}

With the final node-level feature $\mathbf{X}^{(N_\text{\tiny{GC}})} = [\mathbf{x}_1, \dots, \mathbf{x}_{N_I}]^T \in \mathbb{R}^{N_I \times d_{\text{\tiny{GC}}}}$, where $\mathbf{x}_i \in \mathbb{R}^{d_{\text{\tiny{GC}}} \times 1}$ is the final node feature of the $i$-th node $\forall i = 1, \dots, N_I$, the edge-level feature $\boldsymbol{E} \in \mathbb{R}^{N_I \times N_I \times 2 d_{\text{\tiny{GC}}}}$ can be easily obtained via
\begin{equation}
\boldsymbol{E}_{i,j,:} = [\mathbf{x}_i^T, \mathbf{x}_j^T], \quad \forall i,j = 1, \dots, N_I.
\end{equation}

Note that the edge-level feature associated with node self-loops can be expressed as
\begin{equation}
\mathbf{E}_{\mathrm{diag}} = \left\{\boldsymbol{E}_{i,i,:} | i=1,\ldots,N_I \right\}\in \mathbb{R}^{N_I \times 2 d_{\text{\tiny{GC}}}},
\end{equation}
and the edge-level feature associated with node interconnections can be expressed as
\begin{equation}
\mathbf{E}_{\mathrm{low}} = \left\{\boldsymbol{E}_{i,j,:} | i>j, i,j=1,\dots,N_I \right\}
\in \mathbb{R}^{\frac{N_I(N_I-1)}{2} \times 2 d_{\text{\tiny{GC}}}}.
\end{equation}

\subsubsection{Residual Connection}
Graph convolutions generally work well when graphs are not dense. However, during the architecture discovery process, BD-RISs with high circuit complexities correspond to dense graphs, which can lead to insufficient learning in graph convolutions. This is known as oversmoothing caused by repeated neighborhood aggregation due to high graph density and multiple graph convolutional layers\cite{li2018deeper}. To avoid this issue, we design a residual connection to provide additional information when the graphs are dense. Specifically, the residual connection comprises $N_\text{\tiny{RFC}}$ FC layers to generate the intermediate feature $\mathbf{x}_\text{inter} \in \mathbb{R}^{1 \times d_{N_\text{\tiny{RFC}}}^\prime}$ as
\begin{equation}
\mathbf{x}_\text{inter} = \operatorname{ReLU}(\operatorname{ReLU}(\mathbf{x}_{0}^\prime\mathbf{W}_{1}^\prime+\mathbf{b}^\prime_{1})\cdots\mathbf{W}_{N_\text{\tiny{RFC}}}^\prime+\mathbf{b}_{N_\text{\tiny{RFC}}}^\prime),
\end{equation}
where $\mathbf{x}_{0}^\prime = \operatorname{vec}^T(\mathbf{X}_{N}) \in \mathbb{R}^{1 \times 2N_I(N_T+N_R)}$ or $\mathbf{x}_{0}^\prime = \operatorname{vec}^T(\overline{\mathbf{X}}_{N}) \in \mathbb{R}^{1 \times 2N_I(N_T+N_R)}$, $\mathbf{W}_{1}^\prime \in \mathbb{R}^{2N_I(N_T+N_R) \times d_1^\prime}$, $\ldots$, $\mathbf{W}_{N_\text{\tiny{RFC}}}^\prime \in \mathbb{R}^{d_{N_\text{\tiny{RFC}} - 1}^\prime \times d_{N_\text{\tiny{RFC}}}^\prime}$ and $\mathbf{b}_{1}^\prime \in \mathbb{R}^{1 \times d_1^\prime}$, $\ldots$, $\mathbf{b}_{N_\text{\tiny{RFC}}}^\prime \in \mathbb{R}^{1 \times d_{N_\text{\tiny{RFC}}}^\prime}$ represent learnable weight matrices and bias vectors corresponding to the $N_\text{\tiny{RFC}}$ FC layers, respectively.

Then, two linear layers are used to generate residual features for node self-loops and interconnections, respectively, which can be expressed as
\begin{align}
\mathbf{x}_\mathrm{RC, diag} &= (\mathbf{x}_\mathrm{inter}\mathbf{W}_\text{\tiny{RC1}})^T \in \mathbb{R}^{N_I \times 1},\\
\mathbf{x}_\mathrm{RC, low} &= (\mathbf{x}_\mathrm{inter}\mathbf{W}_\text{\tiny{RC2}})^T \in \mathbb{R}^{\frac{N_I(N_I-1)}{2} \times 1},
\end{align}
where $\mathbf{W}_\text{\tiny{RC1}} \in \mathbb{R}^{d_{N_\text{\tiny{RFC}}}^\prime \times N_I}$ and $\mathbf{W}_\text{\tiny{RC2}} \in \mathbb{R}^{d_{N_\text{\tiny{RFC}}}^\prime \times \frac{N_I(N_I-1)}{2}}$ are learnable weight matrices.

\subsubsection{Output Regression}
Once the edge-level features and the residual features are obtained, we combine these two kinds of features and obtain the optimized admittance matrix via output regression layers. The outputs of the performance optimizer vary in the four formulated BD-RIS architecture discovery problems:
\begin{itemize}
\item Ideal BD-RIS: The output susceptance matrix $\mathbf{B}$ will be constructed by combining diagonal elements and off-diagonal elements generated by two regression layers as
\begin{align}
\mathbf{B} &= \mathbf{0},\\
\operatorname{diag}(\mathbf{B}) &= [\mathbf{E}_{\mathrm{diag}}, \mathbf{x}_\mathrm{RC, diag}]\mathbf{W}_{B1} + \mathbf{b}_{B1},\\
\operatorname{tril}(\mathbf{B}) &= [\mathbf{E}_{\mathrm{low}}, \mathbf{x}_\mathrm{RC, low}]\mathbf{W}_{B2} + \mathbf{b}_{B2},\\
\operatorname{triu}(\mathbf{B}) &= (\operatorname{tril}(\mathbf{B}))^T,
\end{align}
where $\mathbf{W}_{B1} \in \mathbb{R}^{(2 d_{\text{\tiny{GC}}} + 1) \times N_I}$, $\mathbf{W}_{B2} \in \mathbb{R}^{(2 d_{\text{\tiny{GC}}} + 1) \times \frac{N_I(N_I - 1)}{2}}$ and $\mathbf{b}_{B1} \in \mathbb{R}^{1 \times N_I}$, $\mathbf{b}_{B2} \in \mathbb{R}^{1 \times \frac{N_I(N_I - 1)}{2}}$ are learnable weight matrices and bias vectors of the two output regression layers for $\mathbf{B}$, respectively.

\item BD-RIS with mutual coupling: The output susceptance matrix $\overline{\mathbf{B}}$ will be constructed by combining diagonal elements and off-diagonal elements generated by two regression layers as
\begin{align}
\overline{\mathbf{B}} &= \mathbf{0},\\
\operatorname{diag}(\overline{\mathbf{B}}) &= [\mathbf{E}_{\mathrm{diag}}, \mathbf{x}_\mathrm{RC, diag}]\mathbf{W}_{\overline{B}1} + \mathbf{b}_{\overline{B}1},\\
\operatorname{tril}(\overline{\mathbf{B}}) &= [\mathbf{E}_{\mathrm{low}}, \mathbf{x}_\mathrm{RC, low}]\mathbf{W}_{\overline{B}2} + \mathbf{b}_{\overline{B}2},\\
\operatorname{triu}(\overline{\mathbf{B}}) &= (\operatorname{tril}(\overline{\mathbf{B}}))^T,
\end{align}
where $\mathbf{W}_{\overline{B}1} \in \mathbb{R}^{(2 d_{\text{\tiny{GC}}} + 1) \times N_I}$, $\mathbf{W}_{\overline{B}2} \in \mathbb{R}^{(2 d_{\text{\tiny{GC}}} + 1) \times \frac{N_I(N_I - 1)}{2}}$ and $\mathbf{b}_{\overline{B}1} \in \mathbb{R}^{1 \times N_I}$, $\mathbf{b}_{\overline{B}2} \in \mathbb{R}^{1 \times \frac{N_I(N_I - 1)}{2}}$ are learnable weight matrices and bias vectors of the two output regression layers for $\overline{\mathbf{B}}$, respectively.

\item Lossy BD-RIS: The output capacitance matrix $\widetilde{\mathbf{C}}$ will be constructed by combining diagonal elements and off-diagonal elements generated by two regression layers as
\begin{align}
\widetilde{\mathbf{C}} &= \mathbf{0},\\
\operatorname{diag}(\widetilde{\mathbf{C}}) &= [\mathbf{E}_{\mathrm{diag}}, \mathbf{x}_\mathrm{RC, diag}]\mathbf{W}_{\widetilde{C}1} + \mathbf{b}_{\widetilde{C}1},\\
\operatorname{tril}(\widetilde{\mathbf{C}}) &= [\mathbf{E}_{\mathrm{low}}, \mathbf{x}_\mathrm{RC, low}]\mathbf{W}_{\widetilde{C}2} + \mathbf{b}_{\widetilde{C}2},\\
\operatorname{triu}(\widetilde{\mathbf{C}}) &= (\operatorname{tril}(\widetilde{\mathbf{C}}))^T,
\end{align}
where $\mathbf{W}_{\widetilde{C}1} \in \mathbb{R}^{(2 d_{\text{\tiny{GC}}} + 1) \times N_I}$, $\mathbf{W}_{\widetilde{C}2} \in \mathbb{R}^{(2 d_{\text{\tiny{GC}}} + 1) \times \frac{N_I(N_I - 1)}{2}}$ and $\mathbf{b}_{\widetilde{C}1} \in \mathbb{R}^{1 \times N_I}$, $\mathbf{b}_{\widetilde{C}2} \in \mathbb{R}^{1 \times \frac{N_I(N_I - 1)}{2}}$ are learnable weight matrices and bias vectors of the two output regression layers for $\widetilde{\mathbf{C}}$, respectively.

\item Discrete-value BD-RIS: Different from previous cases, an $N_b$-bit learnable codebook-based quantizer, i.e., $Q(\cdot)$, is introduced to quantize the diagonal elements and off-diagonal elements generated by two regression layers before they are combined to construct the output susceptance matrix $\underline{\mathbf{B}}$. These can be explicitly expressed as
\begin{align}
\underline{\mathbf{B}} &= \mathbf{0},\\
\underline{\mathbf{B}}_{diag}^\prime &= [\mathbf{E}_{\mathrm{diag}}, \mathbf{x}_\mathrm{RC, diag}]\mathbf{W}_{\underline{B}1} + \mathbf{b}_{\underline{B}1},\\
\underline{\mathbf{B}}_{low}^\prime &= [\mathbf{E}_{\mathrm{low}}, \mathbf{x}_\mathrm{RC, low}]\mathbf{W}_{\underline{B}2} + \mathbf{b}_{\underline{B}2},\\
\operatorname{diag}(\underline{\mathbf{B}}) &= Q(\underline{\mathbf{B}}_{diag}^\prime),\\
\operatorname{tril}(\underline{\mathbf{B}}) &= Q(\underline{\mathbf{B}}_{low}^\prime),\\
\operatorname{triu}(\underline{\mathbf{B}}) &= (\operatorname{tril}(\underline{\mathbf{B}}))^T,
\end{align}
where $\mathbf{W}_{\underline{B}1} \in \mathbb{R}^{(2 d_{\text{\tiny{GC}}} + 1) \times N_I}$, $\mathbf{W}_{\underline{B}2} \in \mathbb{R}^{(2 d_{\text{\tiny{GC}}} + 1) \times \frac{N_I(N_I - 1)}{2}}$ and $\mathbf{b}_{\underline{B}1} \in \mathbb{R}^{1 \times N_I}$, $\mathbf{b}_{\underline{B}2} \in \mathbb{R}^{1 \times \frac{N_I(N_I - 1)}{2}}$ are learnable weight matrices and bias vectors of the two output regression layers for $\underline{\mathbf{B}}$, respectively. To minimize quantization errors, we design a quantizer combining both soft quantization and hard quantization. Specifically, denoting the $N_b$-bit learnable codebook as $\mathcal{B}^{N_b} = \left\{\pm B_{1},\pm B_{2},\ldots,\pm B_{2^{N_b-1}}\right\}$ with $B_{m}>0$, for $m=1,\ldots,2^{N_b-1}$, the quantizer $Q(\cdot)$ determines the best codeword to each element in $\underline{\mathbf{B}}_{diag}^\prime$ and $\underline{\mathbf{B}}_{low}^\prime$ (denoted by $\underline{\mathbf{B}}_{d}^\prime$ for simplicity) during the training phase according to
\begin{align}
q_{d} = \underbrace{\mathcal{B}^{N_b}_{\arg\max\limits_{k} \, \mathbf{D}_{d,k}}}_\text{hard quantization} &+ \underbrace{\sum_{k=1}^K \mathbf{S}_{d,k}\,\mathcal{B}^{N_b}_k}_\text{soft quantization}\notag\\ 
&- \operatorname{stopgrad}\!\left(\sum_{k=1}^K \mathbf{S}_{d,k}\,\mathcal{B}^{N_b}_k\right),\\
\mathbf{D}_{d,k} &= \vert \underline{\mathbf{B}}_{d}^\prime - \mathcal{B}^{N_b}_k \vert,\\
\mathbf{S}_{d,k} &= \frac{\exp\!\left(-\,\mathbf{D}_{d,k}/\tau\right)}
{\sum_{j=1}^K \exp\!\left(-\,\mathbf{D}_{d,j}/\tau\right)},
\end{align}
where $\mathbf{D}$ is the L1 distance matrix reflecting the L1 distances between the elements and codewords, $\mathbf{S}$ is the soft assignment weight matrix for soft quantization, $\tau > 0$ is the temperature parameter controlling the softness of soft quantization, and $\operatorname{stopgrad}(\cdot)$ denotes a gradient stop operation that avoids gradient flow during backpropagation. Note that in the forward pass, only hard quantization takes effect to keep exact quantization based on codewords. While during back propagation, the performance optimizer and the codebook are updated based on the gradient from the soft quantization term and a surrogate gradient $1$ for hard quantization (i.e., $\frac{\partial \mathcal{B}^{N_b}_{\arg\max\limits_{k} \, \mathbf{D}_{d,k}}}{\partial \underline{\mathbf{B}}_{d}^\prime} \approx 1$), making the learning much more effective compared with only hard quantization. 
\end{itemize}

In addition, for the multi-user cases where sum-rate maximization is adopted as the objective function, the framework must perform joint active and passive beamforming design. To achieve this, the downlink precoding matrix $\mathbf{P}$ is dynamically optimized by a dedicated neural network module based on the resultant effective channel $\mathbf{H}_{\text{eff}}$. Specifically, we first extract the real and imaginary parts of $\mathbf{H}_{\text{eff}}$ and flatten them to construct the real-valued input feature vector $\mathbf{x}_{\text{Heff}}$ as
\begin{equation}
\mathbf{x}_{\text{Heff}}  \in \mathbb{R}^{1 \times 2 N_T N_R} = \operatorname{vec}^T([\operatorname{vec}(\Re(\mathbf{H}_{\text{eff}})),\operatorname{vec}(\Im(\mathbf{H}_{\text{eff}}))],
\end{equation}
which is then passed through $N_\text{\tiny{PFC}}$ FC layers with $\operatorname{ReLU}$ activation functions to yield the unnormalized precoding feature vector $\mathbf{p}_{\text{out}}  \in \mathbb{R}^{1 \times 2 N_T N_S}$ as
\begin{equation}
\mathbf{p}_{\text{out}} = \operatorname{ReLU}((\mathbf{x}_{\text{Heff}}\mathbf{W}_{1}^\backprime+\mathbf{b}^\backprime_{1})\cdots)\mathbf{W}_{N_\text{\tiny{PFC}}}^\backprime+\mathbf{b}_{N_\text{\tiny{PFC}}}^\backprime,
\end{equation}
where $\mathbf{W}_{1}^\backprime \in \mathbb{R}^{2 N_T N_R \times d_1^\backprime}$, $\ldots$, $\mathbf{W}_{N_\text{\tiny{PFC}}}^\backprime \in \mathbb{R}^{d_{N_\text{\tiny{PFC}} - 1}^\backprime \times 2 N_T N_S}$ and $\mathbf{b}_{1}^\backprime \in \mathbb{R}^{1 \times d_1^\backprime}$, $\ldots$, $\mathbf{b}_{N_\text{\tiny{PFC}}}^\backprime \in \mathbb{R}^{1 \times 2 N_T N_S}$ represent learnable weight matrices and bias vectors corresponding to the $N_\text{\tiny{PFC}}$ FC layers, respectively. Next, $\mathbf{p}_{\text{out}}$ is split into two equal halves and reshaped to construct the unnormalized complex precoding matrix $\mathbf{P}^\backprime \in \mathbb{C}^{N_T \times N_S}$
\begin{equation}
\mathbf{P}^\backprime = \operatorname{vec}^{-1}(\mathbf{p}_{\text{out}, \Re}) + j\operatorname{vec}^{-1}(\mathbf{p}_{\text{out}, \Im}),
\end{equation}
where $\mathbf{p}_{\text{out}, \Re}$ and $\mathbf{p}_{\text{out}, \Im}$ represent the first and second halves of $\mathbf{p}_{\text{out}}$, respectively, and $\operatorname{vec}^{-1}(\cdot)$ denotes the inverse vectorization operation. Finally, to strictly satisfy the maximum transmit power constraint $P_T$ at the transmitter, a power normalization layer is applied to yield the final downlink precoding matrix $\mathbf{P}$ as
\begin{equation}
\mathbf{P} = \sqrt{P_T} \frac{\mathbf{P}^\backprime}{\Vert \mathbf{P}^\backprime \Vert_F},
\end{equation}
which is subsequently used alongside the BD-RIS passive beamforming to calculate the sum rate.

\section{Numerical Results}\label{Sec. Sim.}
\subsection{General Simulation Setup}
In the simulations, both single-user and multi-user BD-RIS aided systems are considered. In the SU-SISO system, the transmitter, the BD-RIS, and the user are configured with $N_T=1$, $N_I=64$, and $N_k=1$ antennas ($k=1$), respectively. While in the SU-MISO system, the transmitter, the BD-RIS, and the user are configured with $N_T=8$, $N_I=64$, and $N_k=1$ antennas ($k=1$), respectively. In the MU-MIMO system, the transmitter and each of the $K=4$ users are configured with $N_T=8$ and $N_k=2$ antennas ($k=1,2,\cdots,4$), respectively. The path loss at the reference distance $d_0 = 1~\mathrm{m}$ is set as $c_0 = -30$ dB, and the path loss exponents for the transmitter-RIS, RIS-user, and transmitter-user links are given by $a_{IT} = 2.0$, $a_{RI} = 2.8$, and $a_{RT} = 3.5$, respectively. The corresponding large-scale fading gains are computed as $g_{xy} = c_0 (d_{xy}/d_0)^{-a_{xy}}$, where $d_{xy}$ denotes the distance between nodes $x$ and $y$. Specifically, we set the distances $d_{IT}=50$ m, $d_{RI}=2$ m, and $d_{RT}=52$ m. The Rician fading model is adopted for both the transmitter-RIS and RIS-user channels, with Rician factors $K_{\mathrm{T}}=K_{\mathrm{R}}=1$ dB, respectively. Specifically, the small-scale fading components are generated as
\begin{equation}
\mathbf{H}_{k, I} = \sqrt{g_{RI}}\left( \sqrt{\frac{K_{\mathrm{R}}}{1 + K_{\mathrm{R}}}} \mathbf{H}_{k, I}^{\mathrm{LoS}} + \sqrt{\frac{1}{1 + K_{\mathrm{R}}}} \mathbf{H}_{k, I}^{\mathrm{NLoS}} \right),
\end{equation}
\begin{equation}
\mathbf{H}_{IT} = \sqrt{g_{IT}}\left( \sqrt{\frac{K_{\mathrm{T}}}{1 + K_{\mathrm{T}}}} \mathbf{H}_{IT}^{\mathrm{LoS}} + \sqrt{\frac{1}{1 + K_{\mathrm{T}}}} \mathbf{H}_{IT}^{\mathrm{NLoS}} \right),
\end{equation}
where $\mathbf{H}^{\mathrm{LoS}}$ and $\mathbf{H}^{\mathrm{NLoS}}$ denote the deterministic and Rayleigh fading components, respectively. Each user experiences an independent realization of these channels, resulting in $\mathbf{H}_{RI} \in \mathbb{C}^{N_R \times N_I}$ and $\mathbf{H}_{IT} \in \mathbb{C}^{N_I \times N_T}$. The direct transmitter-user channel $\mathbf{H}_{RT}$ is set to zero to emulate a blocked propagation scenario. We set $Y_0=1/50\text{ S}$. In addition, the BD-RIS with mutual coupling is implemented as a uniform planar array (UPA) of elements located in the $x$-$y$ plane with dimensions $N_x \times N_y = 8 \times \frac{N_I}{8}$ and inter-element distance $d$. The RIS elements are thin wire dipoles parallel to the $y$ axis with length $l = \frac{\lambda}{4}$ and radius $r \ll l$, where $\lambda = \frac{c}{f_{MC}}$, $c \approx 3.0 \times 10^8 \ \text{m/s}$, and $f_{MC}=28 \text{ GHz}$ are the wavelength, the speed of light, and the carrier frequency, respectively. Following \cite[Eq.~(91)]{nerini2024global}, we generate $\mathbf{Z}_{II} = \mathbf{Y}_{II}^{-1}$ to represent the mutual coupling at the BD-RIS with $Z_0 = 50~\Omega$. We adopt the same settings of $\mathbf{H}_{IT}$, $\mathbf{H}_{RI}$, and $\mathbf{H}_{RT}$ for $\mathbf{S}_{IT}$, $\mathbf{S}_{RI}$, and $\mathbf{S}_{RT}$, such that $\mathbf{Y}_{IT}$, $\mathbf{Y}_{RI}$, and $\mathbf{Y}_{RT}$ can be obtained according to \cite[Section V-D]{nerini2024universal}. For lossy BD-RIS modeling, we assume the same parameter configurations as \cite{peng2026lossy}, i.e., $f= 2.4 \text{ GHz}$, $L_1 = 6 \text{ nH}$, $L_2 = 0.7 \text{ nH}$, and $\widetilde{C}_{ij} \in [0.35, 3.20]\text{ pF}, \ \forall i,j = 1, \dots, N_I $. The number of channel realizations is $N=100$. The transmit power at the transmitter is set to $P_T = 20\text{ dBm}$ and the noise power is set to $\sigma^2=-80\text{ dBm}$.

\subsection{Learning Process of the LTTADF}
The two proposed modules in the LTTADF, i.e., the architecture generator and the performance optimizer, jointly learn the BD-RIS architectures by maximizing the objectives (i.e., the channel gain for single-user cases and the sum rate for multi-user cases) given specific circuit complexities in an end-to-end manner. During the forward pass of the learning process, the generator first processes the $N$ channel realizations to output the continuous relaxation of the architecture characterization matrix, which is then fed into the performance optimizer alongside the instantaneous channel realizations. The performance optimizer uses the architecture characterization matrix as a structural mask to output the objective-specific variables (e.g., the susceptance matrix and the precoding matrix). While during the backward pass, the gradients of the loss function (derived from the performance objective $O(\cdot)$) are computed at the output of the optimizer and propagated backward. Since the entire LTTADF is designed to be fully differentiable, the gradients flow seamlessly from the performance optimizer back into the architecture generator, allowing the performance optimizer to directly guide the generator to favor architectures that yield higher channel gains or sum rates. Specifically, in the simulations, for single-user cases, the loss function is formulated as the negative average of the channel gain over a batch of $N_{\text{batch}}$ samples, given by
\begin{equation}
\mathcal{L}_{\text{SU}} = -\frac{1}{N_{\text{batch}}} \sum_{n=1}^{N_{\text{batch}}} \mathsf{G}(\mathsf{H}_{\text{eff}}^{(n)}),
\end{equation}
where $\mathsf{G}(\mathsf{H}_{\text{eff}}^{(n)})$ refers to $G(\mathbf{H}_\text{eff}^{(n)})$, $\overline{G}(\overline{\mathbf{H}}_\text{eff}^{(n)})$, $\widetilde{G}(\widetilde{\mathbf{H}}_\text{eff}^{(n)})$, or $\underline{G}(\underline{\mathbf{H}}_\text{eff}^{(n)})$, which computes the channel gain for the $n$-th channel realization under the ideal, antenna-coupled, lossy, or discrete-value BD-RIS scenario, respectively.

For multi-user cases, the loss function is designed as the negative averaged sum rate over the mini-batch, given by
\begin{equation}
\mathcal{L}_{\text{MU}} = -\frac{1}{N_{\text{batch}}} \sum_{n=1}^{N_{\text{batch}}} \mathsf{R}_{\text{sum}}(\mathsf{P}^{(n)}, \mathsf{H}_\text{eff}^{(n)}),
\end{equation}
where $\mathsf{R}_{\text{sum}}(\mathsf{P}^{(n)}, \mathsf{H}_\text{eff}^{(n)})$ refers to $R_{\text{sum}}(\mathbf{P}^{(n)}, \mathbf{H}_\text{eff}^{(n)})$, $\overline{R}_{\text{sum}}(\overline{\mathbf{P}}^{(n)}, \overline{\mathbf{H}}_\text{eff}^{(n)})$, $\widetilde{R}_{\text{sum}}(\widetilde{\mathbf{P}}^{(n)}, \widetilde{\mathbf{H}}_\text{eff}^{(n)})$, or $\underline{R}_{\text{sum}}(\underline{\mathbf{P}}^{(n)}, \underline{\mathbf{H}}_\text{eff}^{(n)})$, which computes the sum rate for the $n$-th channel realization under the ideal, antenna-coupled, lossy, or discrete-value BD-RIS scenario, respectively.

Note that since the performance objective is highly non-linear and takes generally many iterations to converge, the learning process indeed includes two loops: the inner loop updates the performance optimizer to maximize the performance objective, which lasts for $1000$ iterations, and the outer loop updates the architecture generator, which lasts for $100$ epochs. The learning will stop earlier if there is no performance update for $20$ epochs. The hyper-parameter settings of the proposed LTTADF are shown in Table \ref{Hyper}. We use the Adam optimizer\cite{kingma2014adam} and adopt a cosine annealing schedule for the learning rate in the range of $[10^{-5}, 10^{-3}]$, with $N_{\text{batch}} = 100$.

\begin{table}[htbp]
\centering
\caption{Hyper-parameter settings.}
\label{Hyper}
\resizebox{0.4\columnwidth}{!}{%
\begin{tabular}{c|c}
\hline
\textbf{Hyper-parameter}                                & \textbf{Value} \\ \hline
$N_\text{\tiny{FFC}}$                                   & 4              \\ \hline
$d_1, \ldots, d_{N_\text{\tiny{FFC}}}$                   & 768            \\ \hline
$d_\text{\tiny{NE1}},   d_\text{\tiny{NE2}}$            & 384            \\ \hline
$N_\text{\tiny{GC}}$                                    & 3              \\ \hline
$d_\text{\tiny{GC,1}}, \ldots,   d_\text{\tiny{GC}}$    & 384            \\ \hline
$N_\text{\tiny{RFC}}$                                   & 4              \\ \hline
$d_1^\prime, \ldots,   d_{N_\text{\tiny{RFC}}}^\prime$ & 768            \\ \hline
$N_\text{\tiny{PFC}}$                                   & 8              \\ \hline
$d_1^\backprime, \ldots,   d_{N_\text{\tiny{PFC}} - 1}^\backprime$ & 512            \\ \hline
\end{tabular}
}
\end{table}

\subsection{Effectiveness of the Learning-based Architecture Discovery Framework}

We first evaluate the effectiveness of the proposed architecture discovery framework by validating the consistency of learned architectures and optimal architectures found via analytical derivations for ideal BD-RIS in SU-MISO and MU-MIMO systems.

\begin{figure}[htbp]
\centering
\includegraphics[width=0.8\columnwidth]{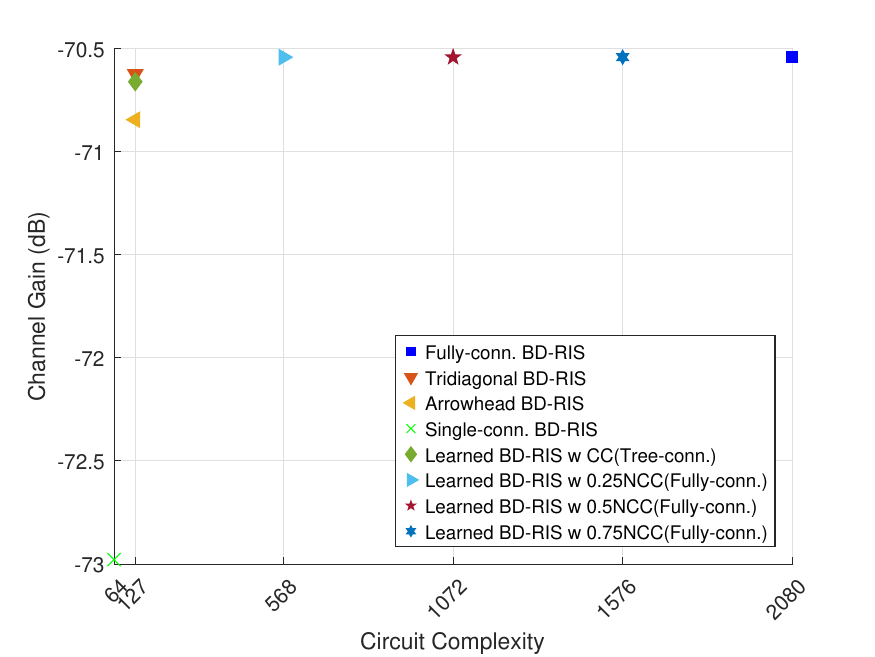}
\caption{Channel gain versus circuit complexity of ideal BD-RIS in an SU-SISO system.}
\label{NI64_SUSISO}
\end{figure}

\begin{figure}[htbp]
\centering
\includegraphics[width=0.8\columnwidth]{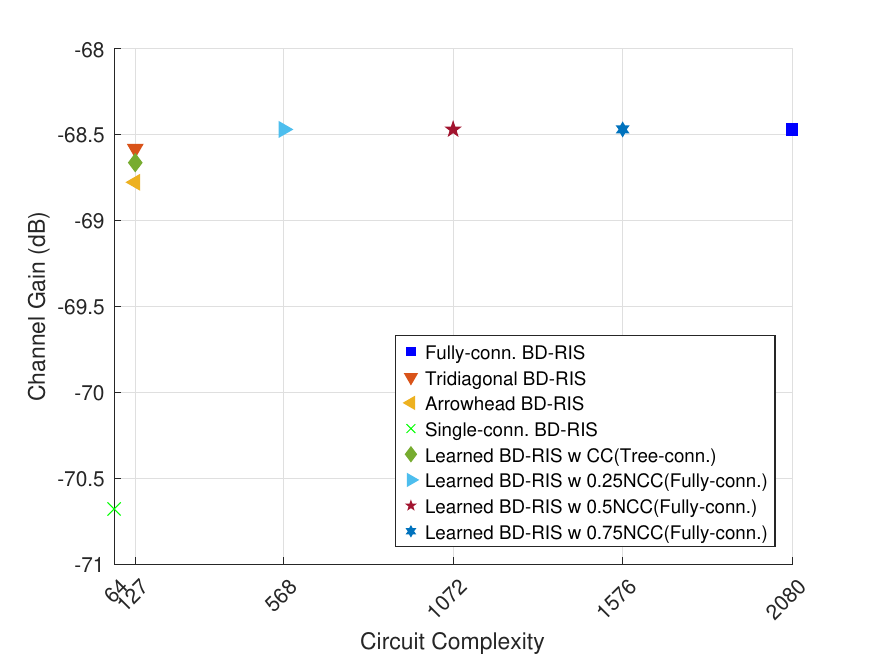}
\caption{Channel gain versus circuit complexity of ideal BD-RIS in an SU-MISO system.}
\label{NI64_SUMISO}
\end{figure}

As shown in Fig. \ref{NI64_SUSISO} and \ref{NI64_SUMISO}, two learned BD-RIS architectures with the circuit complexity equal to $\text{CC(Tree-conn.)}=2 N_I - 1=127$, i.e., the circuit complexity of tree-connected BD-RISs (optimal BD-RIS architectures in SU-SISO and SU-MISO systems\cite{nerini2024diagonal}), demonstrate near-optimal performance approaching the performance of a fully-connected BD-RIS in SU-SISO and SU-MISO systems. In addition, their performance aligns with the performance of two representative tree-connected BD-RISs, i.e., tridiagonal and arrowhead BD-RISs. Note that there are small performance gaps between the performance of the learned architectures and the fully-connected BD-RIS. These performance gaps arise primarily because the LTTADF actually learns an approximation of the mapping between the input channel realizations and optimal solutions through a finite set of learnable parameters, leading to small approximation errors. Such small approximation errors will lead to small gaps between the global optima exactly computed by traditional optimization algorithms and the solutions obtained by the LTTADF. However, these gaps are practically very small and thus can be neglected. In general, the fully-connected BD-RIS will show a slightly better performance due to lower approximation errors, since the fully-connected BD-RIS has higher flexibility for approximation than the others. Similarly, tree-connected BD-RISs theoretically achieve the same optimal performance equal to the fully-connected BD-RIS in SU-SISO/SU-MISO cases\cite{nerini2024diagonal}, while small performance discrepancies between tridiagonal BD-RIS and arrowhead BD-RIS, along with small performance gaps with respect to the fully-connected BD-RIS, can also be observed in Fig. \ref{NI64_SUSISO} and \ref{NI64_SUMISO} due to such approximation errors. Moreover, learned BD-RIS architectures at circuit complexities larger than CC(Tree-conn.) (including architectures with non-diagonal circuit complexities equal to $25\%$, $50\%$, and $75\%$ of the non-diagonal circuit complexity of the fully-connected BD-RIS (denoted by 0.25NCC(Fully-conn.), 0.5NCC(Fully-conn.), and 0.75NCC(Fully-conn.), respectively) show marginal improvements on channel gain compared with BD-RIS architectures with CC(Tree-conn.), indicating that the benefit of increasing circuit complexity becomes very limited when the BD-RIS has a circuit complexity equal to CC(Tree-conn.). These marginal improvements are also due to small approximation errors in the LTTADF, which can be neglected in practice.

\begin{figure}[htbp]
\centering
\includegraphics[width=0.8\columnwidth]{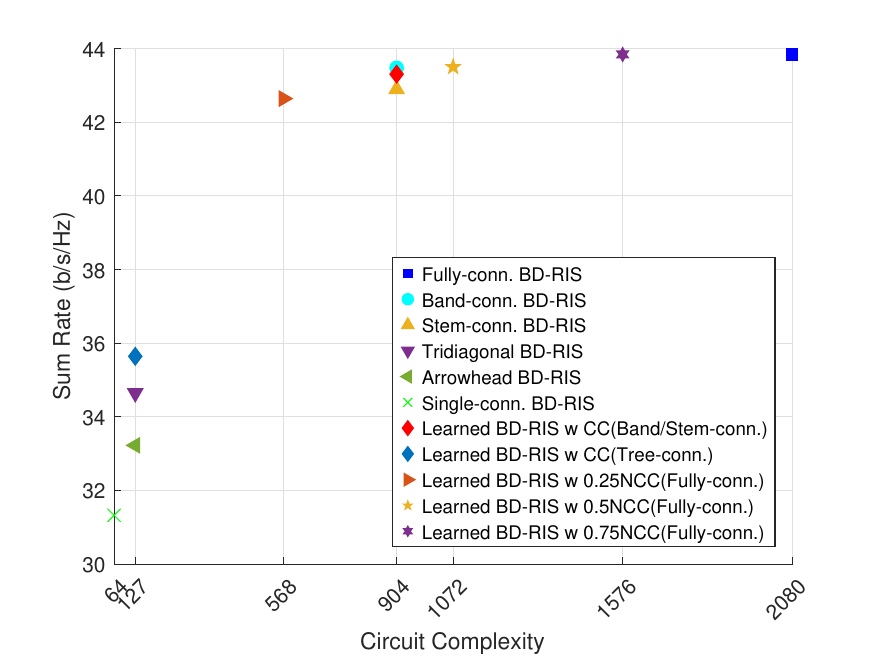}
\caption{Sum rate versus circuit complexity of ideal BD-RIS in an MU-MIMO system.}
\label{SR_NI64_MUMIMO}
\end{figure}

The sum rate achieved by BD-RIS in an MU-MIMO system with different circuit complexities when $N_I=64$ is shown in Fig. \ref{SR_NI64_MUMIMO}. Tree-connected BD-RISs, such as tridiagonal BD-RIS and arrowhead BD-RIS, can theoretically no longer achieve near-optimal performance as the fully-connected BD-RIS in MU-MIMO systems \cite{wu2025beyonddiagonal}, which is verified in Fig. \ref{SR_NI64_MUMIMO}. The learned BD-RIS architecture with CC(Tree-conn.) achieves nearly the same performance as tridiagonal BD-RIS and arrowhead BD-RIS. In addition, the proposed LTTADF can also search over a vast architecture space to discover the most effective BD-RIS architecture with $\text{CC(Band/Stem-conn.)}$ that can achieve near-optimal performance at the same circuit complexity as analytically discovered BD-RIS architectures for MU-MIMO systems (e.g., band-connected and stem-connected BD-RISs\cite{zhou2024novel,wu2025beyonddiagonal}) when BD-RIS is ideal, again demonstrating the effectiveness of the proposed architecture discovery framework. To achieve the optimality, the band/stem width $Q$ characterizing the band-connected and stem-connected BD-RISs must satisfy $Q=2L-1=15$ where $L=\min\{N_R, N_T, \frac{N_I}{2}\}=8$, leading to the circuit complexity $\text{CC(Band/Stem-conn.)}=L(2N_I-2L+1)=904$. The performance of learned BD-RIS architectures with 0.5NCC(Fully-conn.) and 0.75NCC(Fully-conn.) also reveals that increasing the circuit complexity when it is already larger than CC(Band/Stem-conn.) results in negligible performance gain due to approximation errors.

As a brief summary, our results show that the learned BD-RIS architectures align perfectly with those discovered by analytical derivations in the literature under ideal BD-RIS considerations and with the same circuit complexity, demonstrating the effectiveness of the proposed LTTADF. Following this, we employ the LTTADF to discover architectures for BD-RIS with non-idealities to achieve the performance-circuit complexity tradeoff.

\subsection{Architecture Discovery for BD-RIS with Non-Idealities}

\subsubsection{BD-RIS with Mutual Coupling in SU-SISO and MU-MIMO Systems}
\begin{figure}
\centering
\includegraphics[width=0.8\columnwidth]{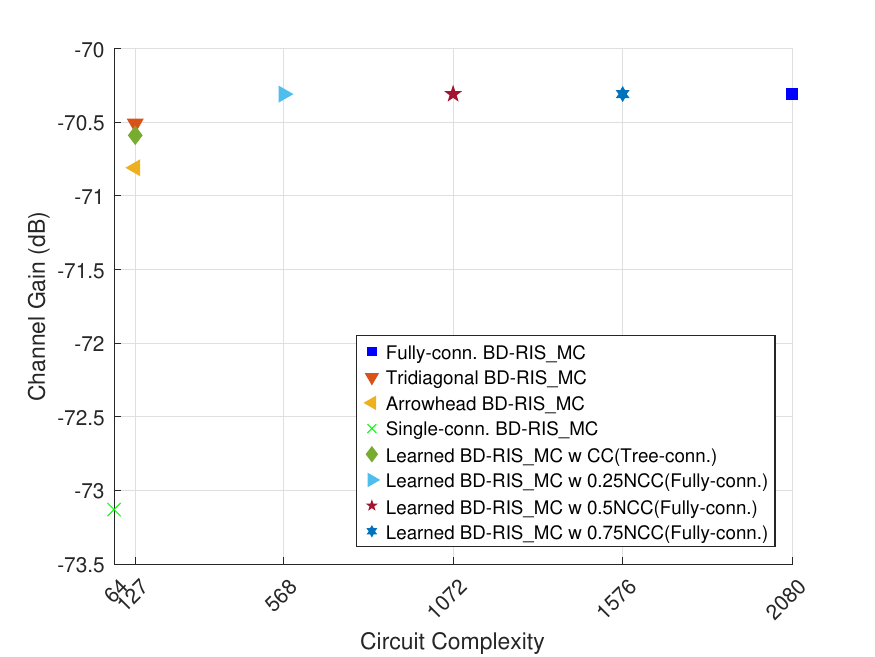}
\caption{Channel gain versus circuit complexity of BD-RIS with mutual coupling and inter-element distance $d=\frac{1}{2}\lambda$ in an SU-SISO system.}
\label{MC_d050_SUSISO}
\end{figure}

The performance of the learned BD-RIS architectures with different circuit complexities when the inter-element distance $d= \frac{1}{2}\lambda$ in an SU-SISO system is shown in Fig. \ref{MC_d050_SUSISO}, where BD-RIS architectures with mutual coupling are denoted by $\text{BD-RIS\_MC}$ for simplicity. The learned BD-RIS architecture with CC(Tree-conn.) is as optimal as fully-connected BD-RIS even when mutual coupling is considered, again supporting that tree-connected BD-RISs, such as tridigonal BD-RIS and arrowhead BD-RIS, are optimal BD-RIS architectures with the lowest circuit complexity\cite{nerini2024global}.

\begin{figure}
\centering
\includegraphics[width=0.8\columnwidth]{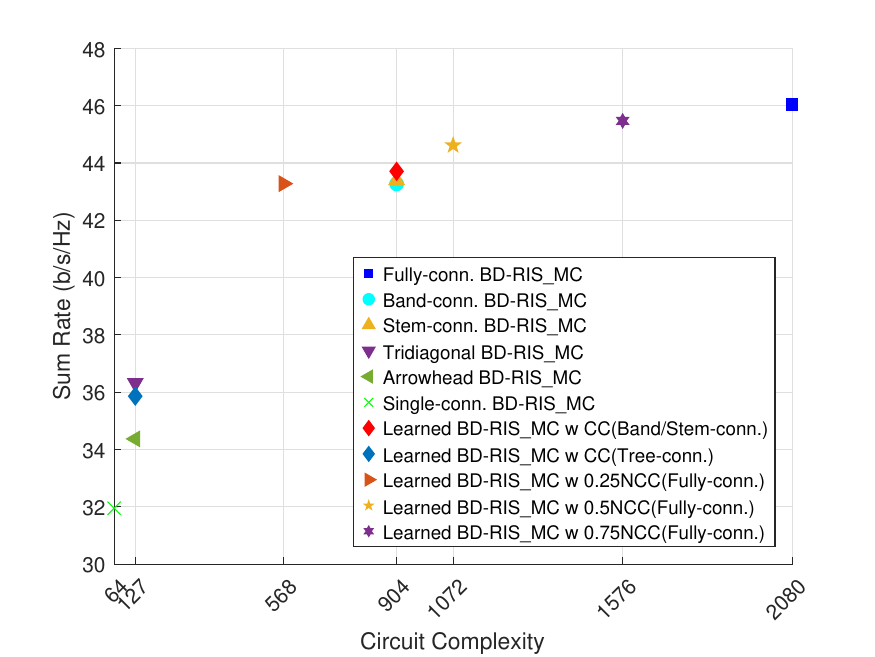}
\caption{Sum rate versus circuit complexity of BD-RIS with mutual coupling and inter-element distance $d=\frac{1}{2}\lambda$ in an MU-MIMO system.}
\label{SR_NI64_MC_d050_MUMIMO}
\end{figure}

As depicted in Fig. \ref{SR_NI64_MC_d050_MUMIMO}, the learned BD-RIS architecture with CC(Band/Stem-conn.), the band-connected BD-RIS, and the stem-connected BD-RIS can achieve near the same performance as the fully-connected BD-RIS in the MU-MIMO system. The results of learned BD-RIS architectures with  0.25NCC(Fully-conn.), 0.5NCC(Fully-conn.), and 0.75NCC(Fully-conn.) show that the performance of BD-RIS with mutual coupling saturates near CC(Band/Stem-conn.). These results reveal that mutual coupling at the BD-RIS will not significantly affect the optimal architecture of BD-RIS in MU-MIMO systems.

\subsubsection{Lossy BD-RIS in SU-SISO and MU-MIMO Systems}
\begin{figure}
\centering
\includegraphics[width=0.8\columnwidth]{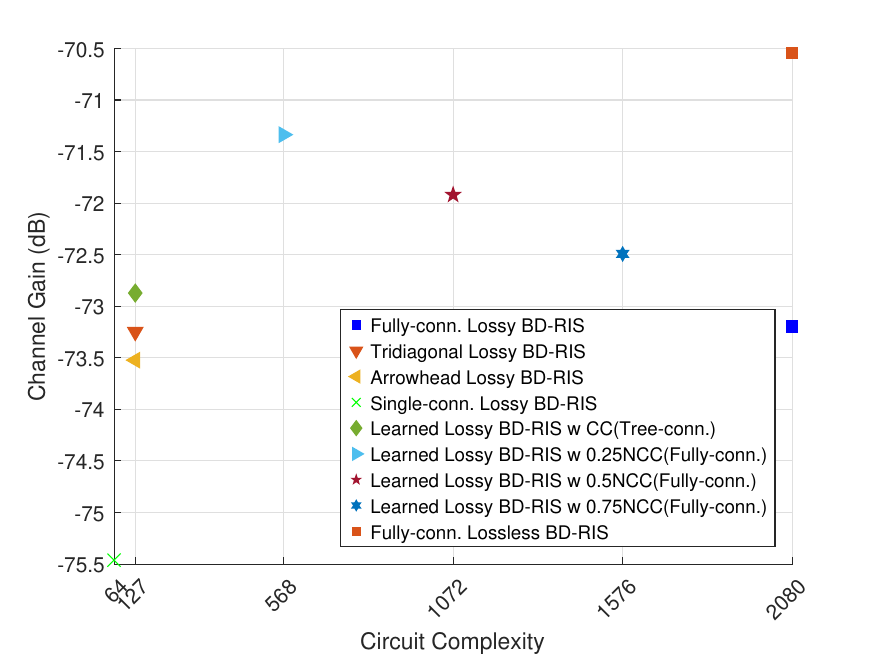}
\caption{Channel gain versus circuit complexity of lossy BD-RIS with $R=1$ in an SU-SISO system.}
\label{NI64_Lossy_SUSISO}
\end{figure}

\begin{figure}
\centering
\includegraphics[width=0.8\columnwidth]{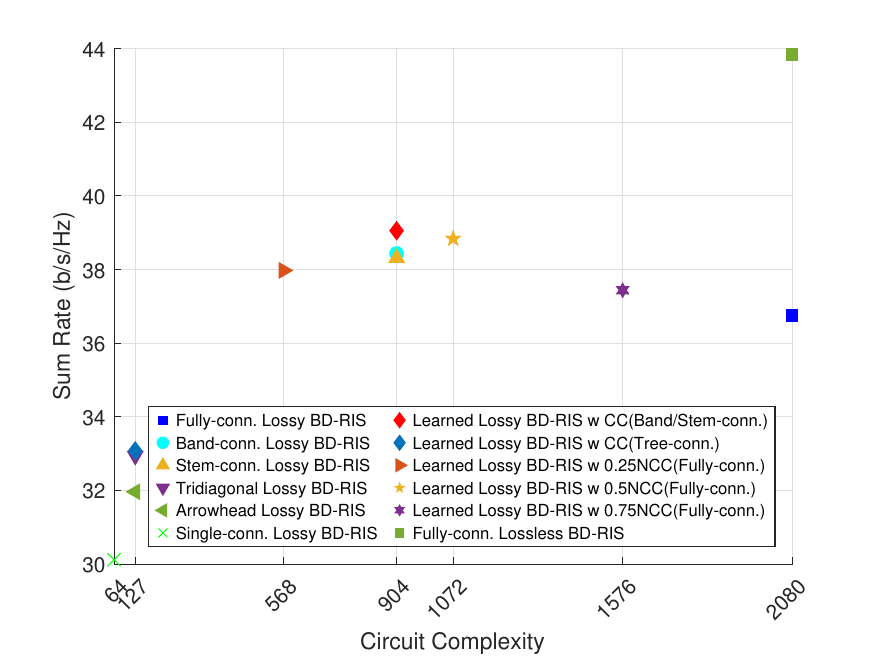}
\caption{Sum rate versus circuit complexity of lossy BD-RIS with $R=1$ in an MU-MIMO system.}
\label{SR_NI64_Lossy_MUMIMO}
\end{figure}

The performance of learned BD-RIS architectures and benchmark BD-RIS architectures versus the circuit complexity in lossy BD-RIS (with $R=1$) assisted SU-SISO and MU-MIMO systems is shown in Fig. \ref{NI64_Lossy_SUSISO} and Fig. \ref{SR_NI64_Lossy_MUMIMO}, respectively. Different from previous results, it can be clearly seen that the fully-connected BD-RIS can no longer achieve the optimal performance. Similarly, tree-connected BD-RISs in the SU-SISO system and band-connected/stem-connected BD-RISs in the MU-MIMO system can neither achieve optimal results, respectively. In the SU-SISO system, the learned lossy BD-RIS architecture with 0.25NCC(Fully-conn.) achieves the best performance compared with the others. While in the MU-MIMO system, the learned lossy BD-RIS architecture with CC(Band/Stem-conn.) achieves the best performance. Comparing Fig. \ref{NI64_Lossy_SUSISO} and Fig. \ref{SR_NI64_Lossy_MUMIMO} with Fig. \ref{NI64_SUSISO} and Fig. \ref{SR_NI64_MUMIMO}, it is clear that, in the lossy case, the performance–circuit complexity relationship exhibits concavity with a decline in performance, in contrast to the increase with saturation in performance observed in the ideal case. The physical mechanism stems from a fundamental trade-off between the beamforming gain of the reconfigurable impedance network and the insertion loss inherent to each tunable admittance component. At lower circuit complexities, the enhanced wave manipulation flexibility provided by added interconnections yields a beamforming gain that easily outweighs the accumulated losses. However, as connectivity scales toward a fully-connected architecture, the marginal beamforming gain diminishes while aggregated resistive losses continue to increase. Consequently, the concave trends in Fig. \ref{NI64_Lossy_SUSISO} and Fig. \ref{SR_NI64_Lossy_MUMIMO} capture the exact tipping point where the insertion losses of a dense admittance network eclipse its beamforming benefits. These results provide insightful guidance to achieve the tradeoff between the performance and circuit complexity of lossy BD-RIS: increasing circuit complexity can be detrimental for lossy BD-RIS, so that learning the optimal architecture for lossy BD-RIS is not only meaningful to reduce the circuit complexity but also to avoid negative effects caused by losses on BD-RIS performance.

\subsubsection{Discrete-Value BD-RIS in SU-SISO and MU-MIMO Systems}
\begin{figure}
\centering
\includegraphics[width=0.8\columnwidth]{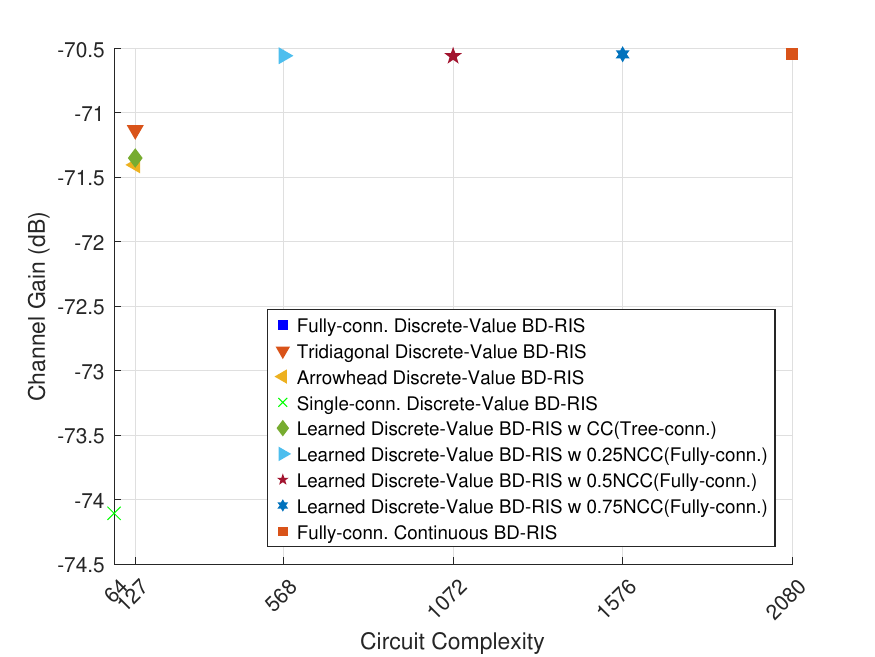}
\caption{Channel gain versus circuit complexity of discrete-value BD-RIS with $N_b=4$ in an SU-SISO system.}
\label{NI64_B4_SUSISO}
\end{figure}

\begin{figure}
\centering
\includegraphics[width=0.8\columnwidth]{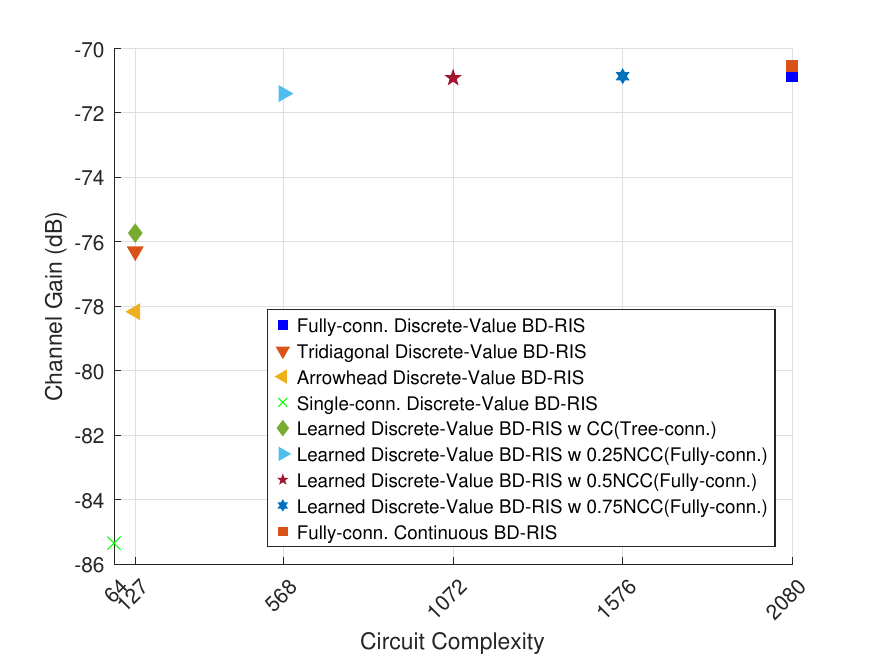}
\caption{Channel gain versus circuit complexity of discrete-value BD-RIS with $N_b=1$ in an SU-SISO system.}
\label{NI64_B1_SUSISO}
\end{figure}

In Fig. \ref{NI64_B4_SUSISO} and Fig. \ref{NI64_B1_SUSISO}, we present the performance of discrete-value BD-RIS in an SU-SISO system with $N_b=4$ and $N_b=1$, respectively. The temperature parameter controlling the softness of soft quantization is set as $\tau = 0.1$. It can be observed that tree-connected BD-RISs are no longer optimal architectures in both quantization bit settings, while the performance of discrete-value BD-RIS saturates as the circuit complexity increases after 0.25NCC(Fully-conn.) when $N_b=4$ and after 0.5NCC(Fully-conn.) when $N_b=1$, respectively. These indicate that, limited by quantization bits, it is necessary to increase the circuit complexity of discrete-value BD-RIS to compensate for quantization errors and performance loss. On the contrary, for BD-RISs with higher circuit complexities, fewer quantization bits can also guarantee near-optimal performance. For example, as shown in Fig. \ref{NI64_B1_SUSISO}, even using $1$ quantization bit for discrete-value fully-connected BD-RIS will only cause a small performance drop compared with continuous fully-connected BD-RIS. These reflect the tradeoffs among the performance, the circuit complexity, and the number of quantization bits of discrete-value BD-RIS in an SU-SISO system.

\begin{figure}
\centering
\includegraphics[width=0.8\columnwidth]{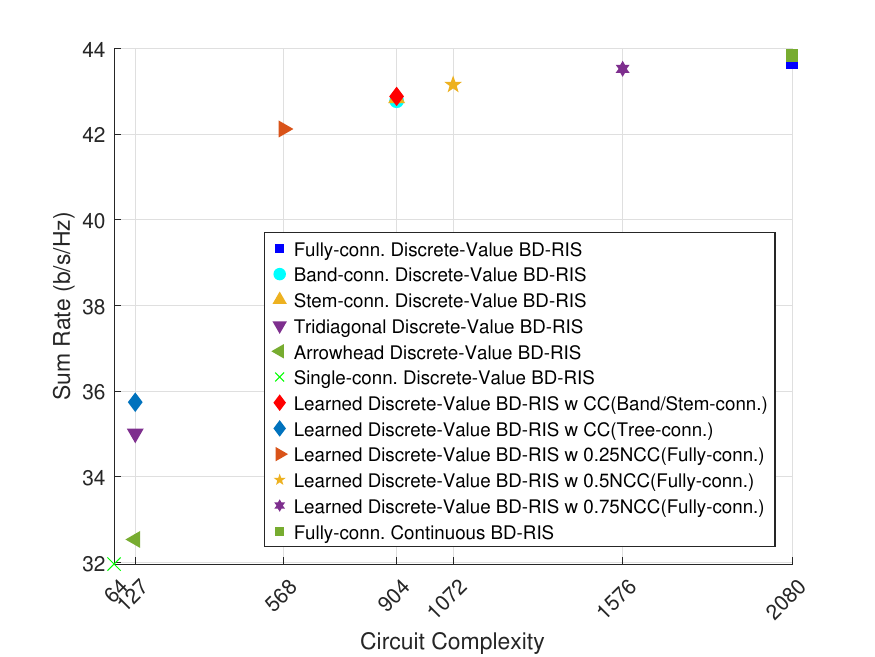}
\caption{Sum rate versus circuit complexity of discrete-value BD-RIS with $N_b=4$ in an MU-MIMO system.}
\label{SR_NI64_B4_MUMIMO}
\end{figure}

\begin{figure}
\centering
\includegraphics[width=0.8\columnwidth]{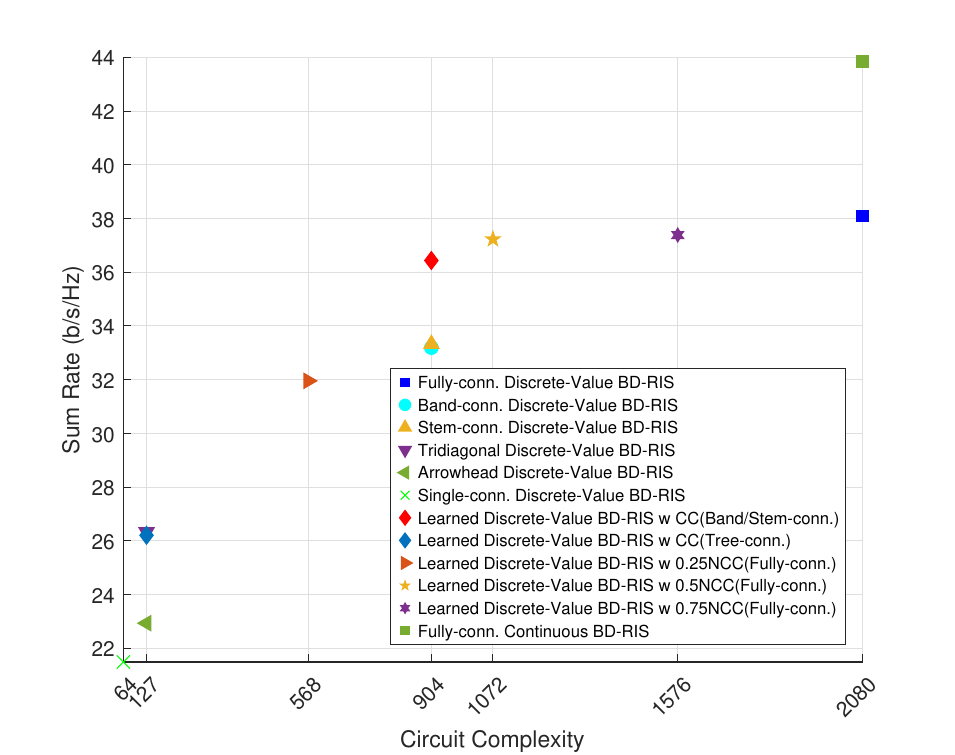}
\caption{Sum rate versus circuit complexity of discrete-value BD-RIS with $N_b=1$ in an MU-MIMO system.}
\label{SR_NI64_B1_MUMIMO}
\end{figure}

The performance of learned discrete-value BD-RIS architectures and benchmark BD-RIS architectures in an MU-MIMO system with $N_b=4$ and $N_b=1$ is shown in Fig. \ref{SR_NI64_B4_MUMIMO} and Fig. \ref{SR_NI64_B1_MUMIMO}, respectively. It can be observed that band-connected/stem-connected BD-RISs are no longer optimal architectures in both quantization bit settings. Similar to the SU-SISO case, the tradeoffs among the performance, the circuit complexity, and the number of quantization bits also exist in MU-MIMO systems. However, unlike the single-user case, the multi-user system demands more quantization bits, where a 1-bit discrete-value fully-connected BD-RIS suffers a significant performance loss compared to the continuous fully-connected BD-RIS. Not surprisingly, the 1-bit discrete-value BD-RIS requires a much higher circuit complexity to achieve the same performance as the fully-connected discrete-value BD-RIS in the MU-MIMO system.

\begin{figure}[htbp]
\centering
\includegraphics[width=0.8\columnwidth]{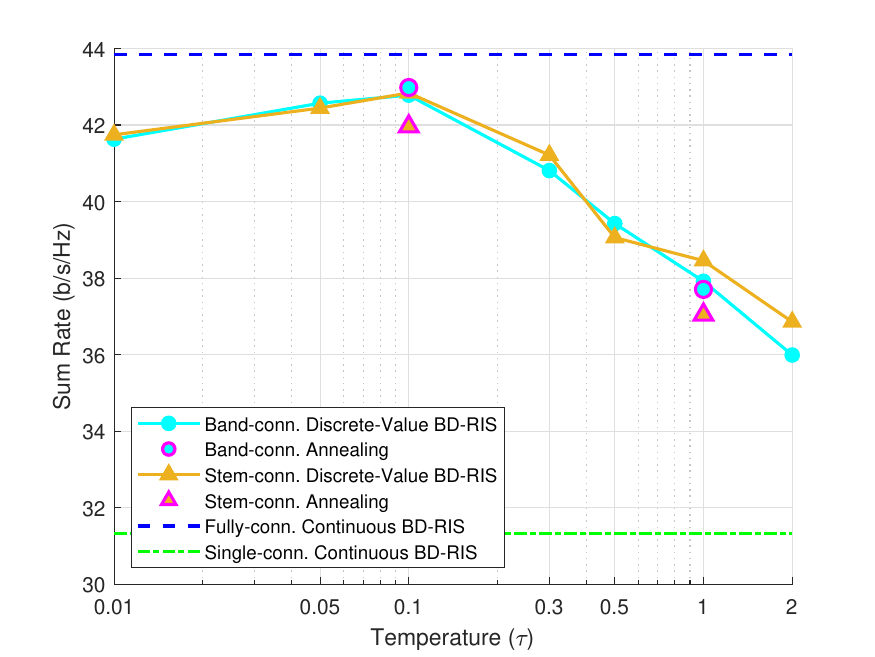}
\caption{Sum rate versus the temperature parameter $\tau$ of discrete-value BD-RIS with $N_b=4$ in an MU-MIMO system.}
\label{SR_NI64_B4_MUMIMO_Tau}
\end{figure}

Since the temperature parameter $\tau$ is important to bridge the continuous relaxation and the hard discrete solution during backpropagation, we further conducted an extensive sensitivity analysis to numerically evaluate the impact of different $\tau$ values, as well as the annealing schedule for $\tau$, on the final discrete solution of the proposed framework for band-connected and stem-connected discrete-value BD-RIS. Specifically, we evaluated fixed $\tau$ values of $[0.01, 0.05, 0.1, 0.3, 0.5, 1.0, 2.0]$ and two cosine annealing schedules decaying from $1.0$ / $0.1$ to $0.0$, respectively. The results shown in Fig. \ref{SR_NI64_B4_MUMIMO_Tau} demonstrate that a fixed $\tau = 0.1$ strikes an effective balance, providing sufficient gradient flow while maintaining a tight bound to the hard discrete solution. While an annealing schedule induces higher hyperparameter tuning complexity, it does not guarantee consistent and significant performance superiority over the fixed $\tau = 0.1$ setting. However, it is worth emphasizing that $\tau$ values in $[0.01, 0.3]$, as well as a cosine annealing schedule with an appropriate initial setting for $\tau$, maintain acceptable sum rate performance, showing that the proposed framework exhibits strong robustness to the selection of $\tau$.

\subsubsection{Robustness of the LTTADF in Asymmetric MU-MIMO Networks}

\begin{figure}[htbp]
\centering
\includegraphics[width=0.8\columnwidth]{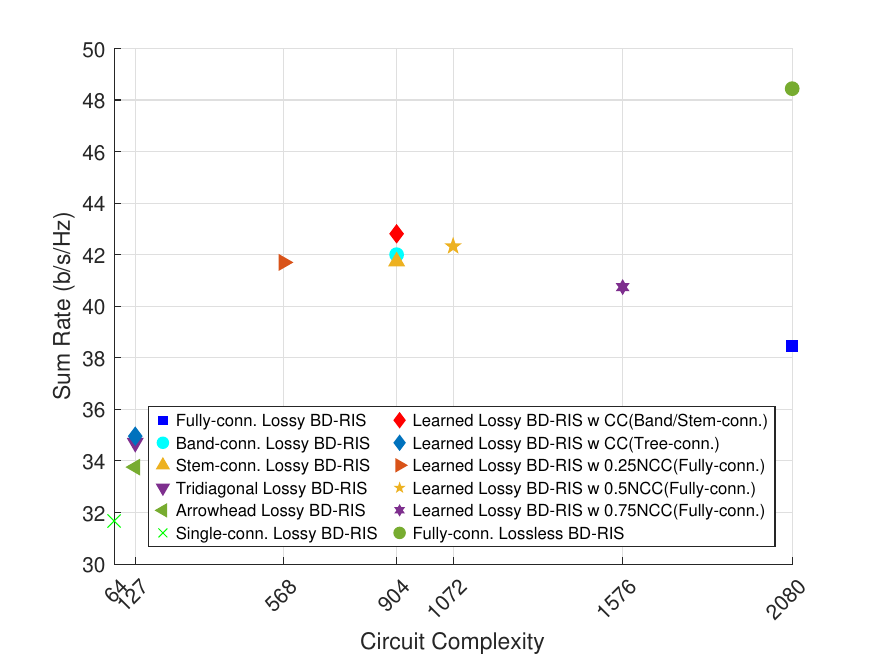}
\caption{Sum rate versus circuit complexity of lossy BD-RIS with a heterogeneous user antenna setup ($4$ users with $N_k \in [2, 3, 1, 4]$) and $R=1$ in an MU-MIMO system.}
\label{SR_NI64_Lossy_MUMIMO_AsymNk}
\end{figure}

To further validate the robustness of the proposed framework under asymmetric network conditions, we evaluate the sum rate performance of lossy BD-RIS in an MU-MIMO system with a heterogeneous user antenna setup as an example, where the number of antennas for $K=4$ users is configured as $N_k \in \{2, 3, 1, 4\}$ and the transmitter antenna number remains $N_T=8$. As depicted in Fig. \ref{SR_NI64_Lossy_MUMIMO_AsymNk}, the fundamental tradeoff between beamforming gain and insertion loss remains highly consistent with the observations in the symmetric case (i.e., Fig. \ref{SR_NI64_Lossy_MUMIMO}). The sum rate exhibits the same concave trend concerning circuit complexity. Specifically, the fully-connected lossy BD-RIS still suffers from significant performance degradation due to the accumulation of resistive losses at maximum circuit complexity. In contrast, the learned lossy BD-RIS architecture with CC(Band/Stem-conn.) successfully captures the optimal tipping point, achieving the peak sum rate among all lossy configurations. These results confirm that the proposed architecture learning strategy effectively generalizes to heterogeneous MU-MIMO systems, consistently avoiding the detrimental hardware losses of over-parameterized networks while maximizing achievable performance.

\subsection{Computational Overhead and Physical Implementation Considerations}

When evaluating the computational overhead and practical deployability of the proposed LTTADF, it is crucial to distinguish between the offline hardware design phase and the online deployment phase. The primary focus of this work is the offline discovery of the optimal static hardware architecture $\mathbf{A}$ over a statistical distribution of channels. Since the architecture is fixed at manufacturing to avoid the prohibitive costs and high latency of real-time switching networks, architecture discovery is exclusively an offline procedure. Consequently, the training time of the LTTADF does not impose a bottleneck on practical deployment. For online deployment, the system only needs to adjust the tunable admittance components based on instantaneous CSI. For this real-time task, the GNN-based performance optimizer offers significant time-efficiency advantages over traditional iterative algorithms, such as the block coordinate descent (BCD) algorithm \cite{peng2026lossy}. While conventional methods often require hundreds of iterations involving complex matrix inversions, the proposed performance optimizer computes the solution via a single forward pass\footnote{It should be noted that achieving robust generalization for such single forward-pass inference requires extensively training the GNN-based performance optimizer over a massive dataset of channel realizations once the hardware architecture $\mathbf{A}$ is fixed.}, which relies entirely on deterministic, highly parallelizable matrix multiplications. Even in cases where a few gradient-based fine-tuning iterations are employed to further mitigate severe non-idealities, the inference latency remains significantly faster than traditional optimization algorithms, ensuring that the online beamforming design can be executed within the channel coherence interval.

\begin{figure*}[htbp]
    \centering

    \begin{subfigure}[t]{0.25\linewidth}
        \centering
        \includegraphics[width=\linewidth]{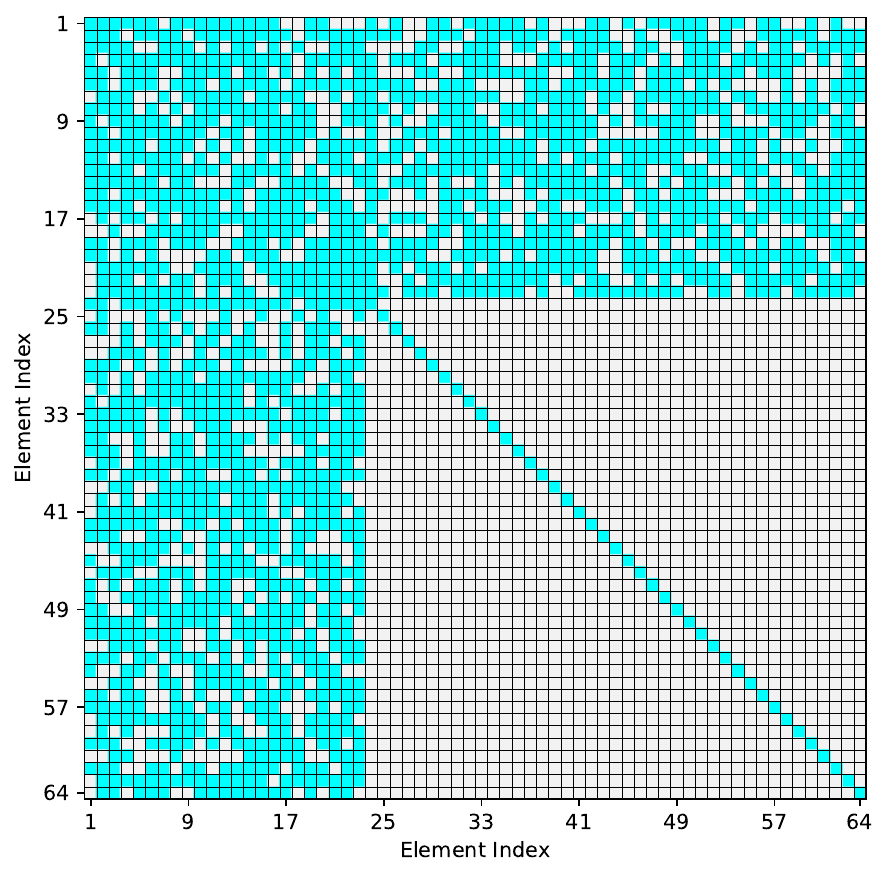}
        \caption{Learned BD-RIS with mutual coupling architecture with CC(Band/Stem-conn.) in Fig. \ref{SR_NI64_MC_d050_MUMIMO}.}
        \label{fig:subfig1}
    \end{subfigure}
    \hfill
    \begin{subfigure}[t]{0.25\linewidth}
        \centering
        \includegraphics[width=\linewidth]{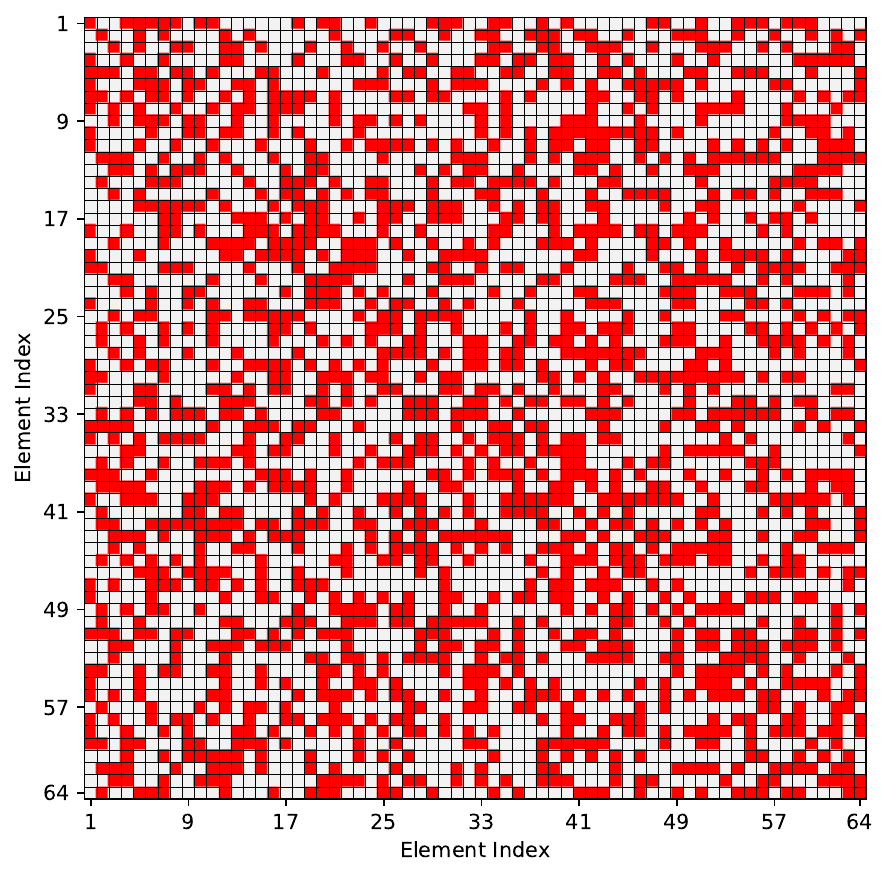}
        \caption{Learned lossy BD-RIS architecture with CC(Band/Stem-conn.) in Fig. \ref{SR_NI64_Lossy_MUMIMO}.}
        \label{fig:subfig2}
    \end{subfigure}
    \hfill
    \begin{subfigure}[t]{0.25\linewidth}
        \centering
        \includegraphics[width=\linewidth]{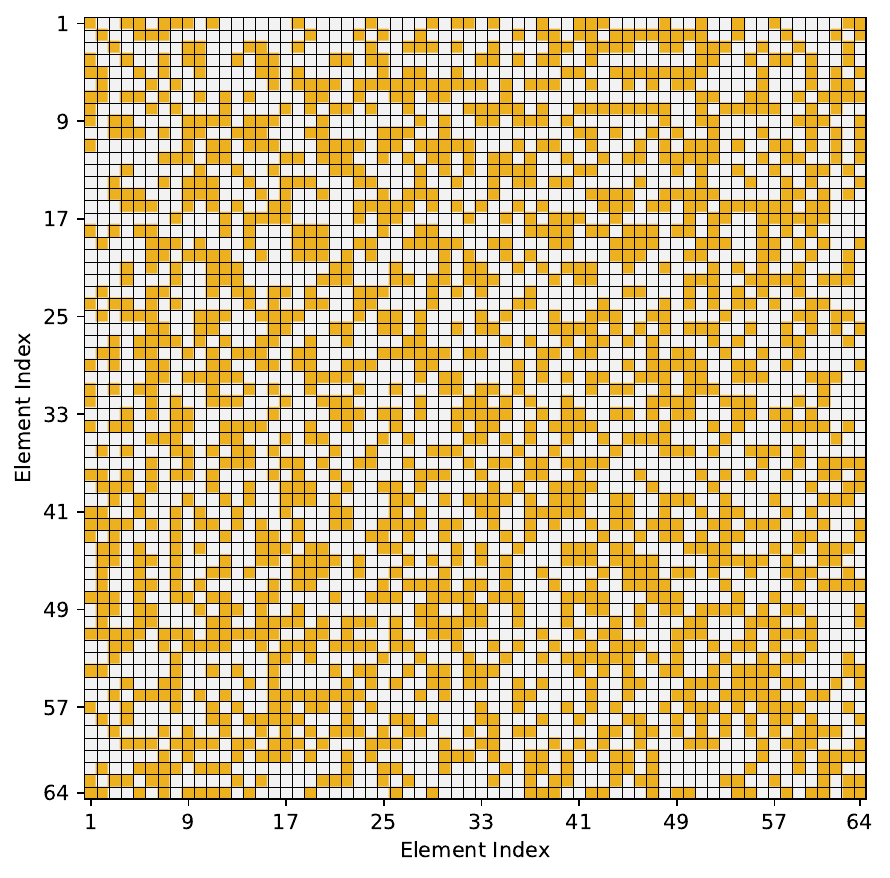}
        \caption{Learned discrete-value BD-RIS architecture with CC(Band/Stem-conn.) in Fig. \ref{SR_NI64_B4_MUMIMO}.}
        \label{fig:subfig3}
    \end{subfigure}
    
    \caption{Visualization of learned BD-RIS architectures.}
    \label{arch_visualization}

\end{figure*}

It is also important to contextualize the aforementioned architecture discovery within the practical constraints of physical hardware implementation. To provide some physical intuitions, Fig. \ref{arch_visualization} visualizes three BD-RIS architectures obtained by the LTTADF under three BD-RIS non-ideal scenarios, where connections between two elements are marked with colors. In our graph-based abstraction, connections between arbitrary elements are theoretically permitted. As can be observed from Fig. \ref{arch_visualization}, these learned architectures involve connections between some distant elements. However, in a physical printed circuit board (PCB) implementation, connecting distant elements necessitates long transmission lines, which may naturally introduce distance-dependent transmission line effects, such as Ohmic losses, group delay, and crosstalk, and thus render refined PCB routing techniques during hardware layout indispensable. While the explicit modeling of these specific layout-dependent dynamics falls beyond the scope of our current formulations, the proposed framework successfully captures the primary hardware constraints of BD-RIS. Consequently, the architectural insights and optimization strategies presented in this paper establish a crucial foundational framework for practical BD-RIS implementation and deployment. Moving forward, integrating rigorous transmission line theory to explicitly capture these spatial routing non-idealities during the learning process is a highly complex task. We reserve this advanced physical modeling for future work to further bridge the gap between graph-theoretic architecture discovery and practical microwave engineering.

\section{Conclusion} \label{Sec. Con.}
This paper investigates the architecture discovery for BD-RIS with non-idealities using machine learning. We propose a learning-based architecture discovery framework, namely the LTTADF, to discover optimal architectures for BD-RIS with mutual coupling, lossy BD-RIS, and discrete-value BD-RIS with given circuit complexities. The LTTADF is applicable to both single-user and multi-user cases under all antenna settings (including SISO, MISO, MIMO, etc.). In the simulations, we first compare learned BD-RIS architectures with tree-connected BD-RISs in SU-SISO and SU-MISO systems and with band-connected/stem-connected BD-RISs for ideal BD-RISs, and then compare them with tree-connected BD-RISs in SU-SISO systems for BD-RISs with mutual coupling. Simulation results show that the learned BD-RIS architectures align perfectly with those analytically discovered BD-RIS architectures in the literature, demonstrating the effectiveness of the LTTADF. We then employ the LTTADF to discover optimal architectures for BD-RIS with non-idealities, which provides meaningful insights for achieving the performance-circuit complexity tradeoff in the presence of non-idealities. Specifically, mutual coupling at the BD-RIS does not affect the optimal BD-RIS architecture in MU-MIMO systems. For lossy BD-RIS, increasing circuit complexity can be detrimental, making it important to learn an optimal architecture that minimizes complexity while mitigating loss effects. Furthermore, results based on discrete-value BD-RIS reveal a tradeoff between circuit complexity and quantization resolution, where one can compensate for the limitations of the other.

\bibliographystyle{IEEEtran}
\bibliography{text_final}

\begin{IEEEbiography}
    [{\includegraphics[width=1in,height=1.25in,clip,keepaspectratio]{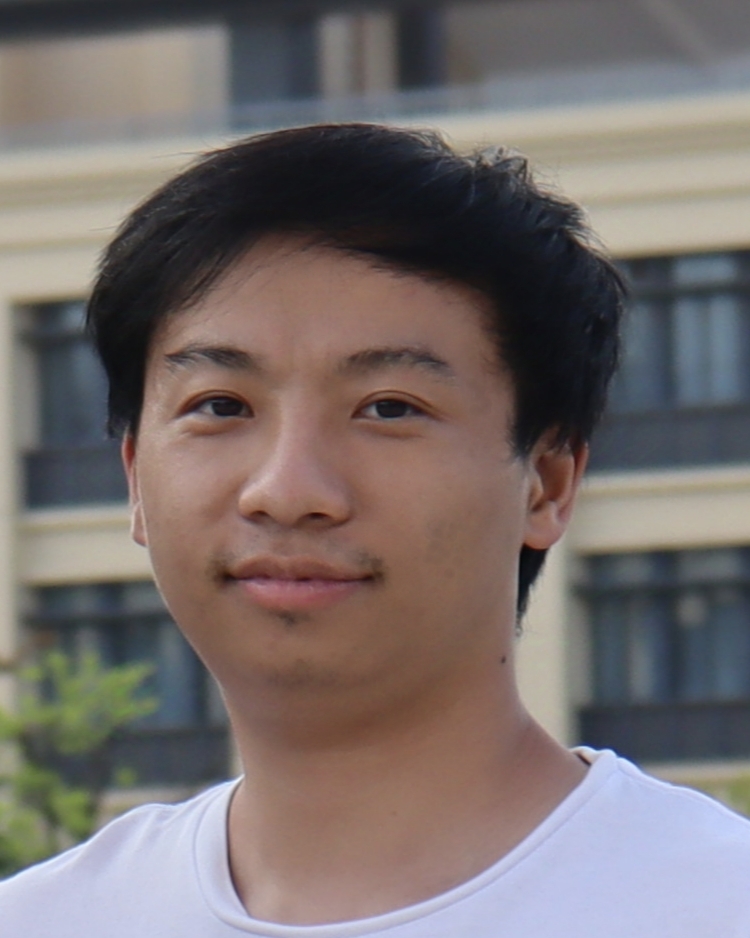}}]{Binggui Zhou} (Member, IEEE) is currently a Postdoctoral Research Associate with the Department of Electrical and Electronic Engineering, Imperial College London, London, U.K. He received his B.Eng. degree in Electrical Engineering from Jinan University, Zhuhai, China, in 2018, and his M.Sc. degree and Ph.D. degree in Electrical and Computer Engineering from the University of Macau, Macao SAR, China, in 2021 and 2024, respectively. His research interests lie in artificial intelligence (AI) and AI-native wireless systems, with a focus on massive MIMO, reconfigurable intelligent surface (RIS), and integrated sensing and communications (ISAC). He serves as an Associate Editor for IEEE WCL. He has served as a General Co-Chair for workshops in IEEE GLOBECOM 2026 and IEEE/CIC ICCC 2026, and as a Technical Program Committee (TPC) member for several flagship international conferences (e.g., IEEE GLOBECOM, IEEE ICC, and IEEE VTC). He was a recipient of the European Union's Marie Skłodowska-Curie Actions (MSCA) Postdoctoral Fellowship.
\end{IEEEbiography}

\begin{IEEEbiography}
    [{\includegraphics[width=1in,height=1.25in,clip,keepaspectratio]{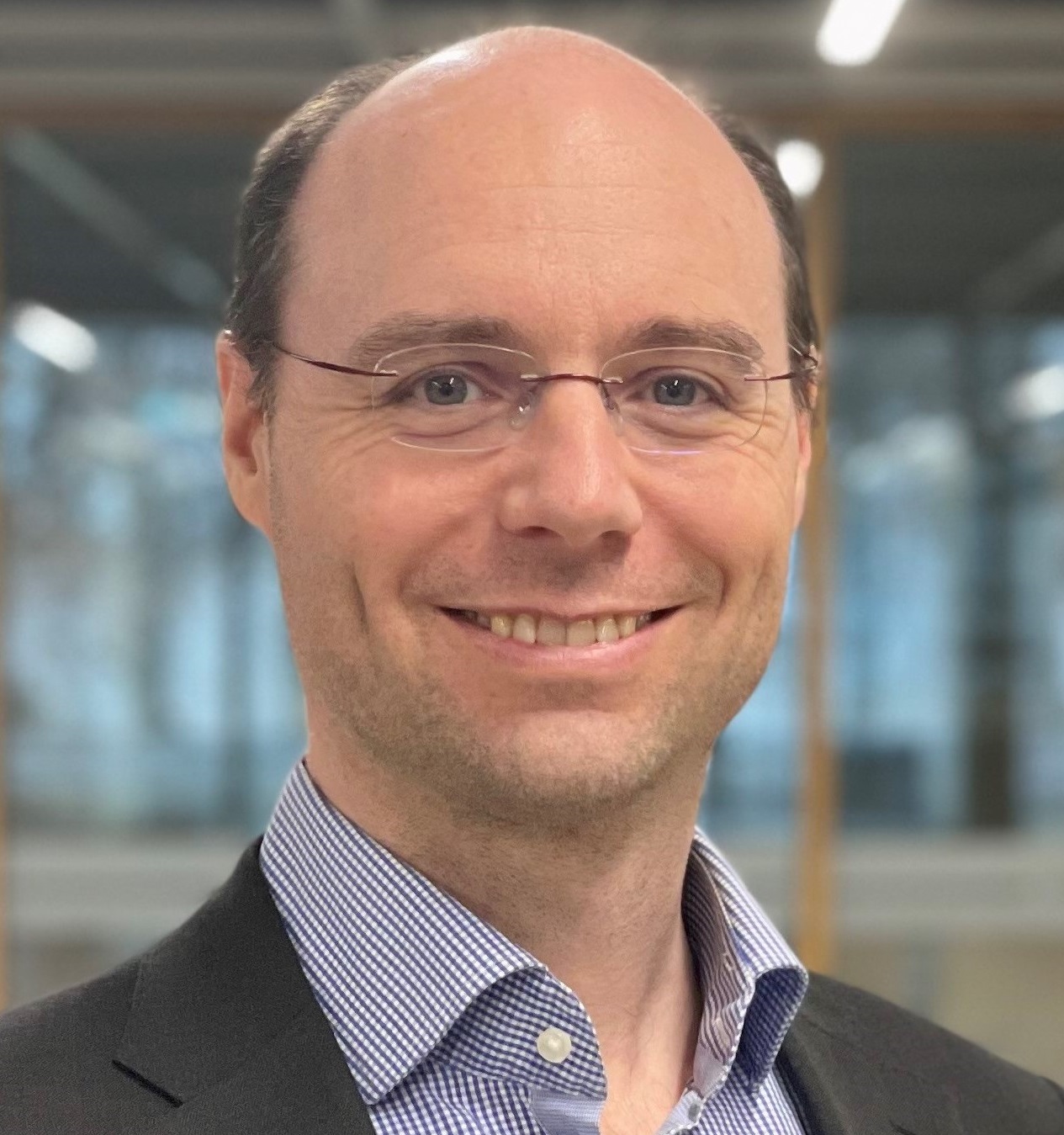}}]{{Bruno Clerckx}} (Fellow, IEEE) received the MSc and Ph.D. degrees in Electrical Engineering from Université Catholique de Louvain, Belgium, and the Doctor of Science (DSc) degree from Imperial College London, U.K. He spent many years in industry with Silicon Austria Labs (SAL), Austria, where he was the Chief Technology Officer (CTO) responsible for all research areas of Austria's top research center for electronic based systems and with Samsung Electronics, South Korea, where he actively contributed to 4G (3GPP LTE/LTE-A and IEEE 802.16m). He is currently a Professor and the Head of the Communications and Signal Processing Group within the Electrical and Electronic Engineering Department, Imperial College London, London, U.K. He has authored two books on “MIMO Wireless Communications” and “MIMO Wireless Networks”, over 350 peer-reviewed international research papers, and 150 standards contributions, and is the inventor of 80 issued or pending patents among which several have been adopted in the specifications of 4G standards and are used by billions of devices worldwide. His research spans the general area of wireless communications and signal processing for wireless networks. He received the prestigious Blondel Medal 2021 from France for exceptional work contributing to the progress of Science and Electrical and Electronic Industries, the 2022 Adolphe Wetrems Prize in mathematical and physical sciences and the 2025 Georges Vanderlinden Prize in Electromagnetism and Telecommunications from Royal Academy of Belgium, multiple awards from Samsung, IEEE best student paper award, IEEE Globecom 2025 best paper award, EURASIP (European Association for Signal Processing) best paper award 2022, and 2026 IEEE Marconi Paper Award in Wireless Communications. He is the recipient of an ERC Advanced Grant.
\end{IEEEbiography}

\end{document}